\documentstyle[10pt,aaspp4]{article}

\newcommand{\bq}{\begin{equation}}
\newcommand{\eq}{\end{equation}}

\begin{document}
\def\refitem{\par\parskip 0pt\noindent\hangindent 20pt}
\normalsize
 
\title{Observational Evidence from Supernovae for an Accelerating Universe and a Cosmological Constant \\
\vspace*{1.0cm}
{\it To Appear in the Astronomical Journal}}
\vspace*{0.3cm}

Adam G. Riess\footnote{Department of Astronomy, University of California, 
Berkeley, CA 94720-3411}, Alexei V. Filippenko$^1$,
Peter Challis\footnote{Harvard-Smithsonian Center for Astrophysics, 60
Garden St., Cambridge, MA 02138},
 Alejandro Clocchiatti\footnote{Departamento de Astronom\'{\i}a y Astrof\'{\i}sica
Pontificia Universidad Cat\'olica, Casilla 104, Santiago 22, Chile},
Alan Diercks\footnote{Department of Astronomy, University of Washington, Seattle, WA 
98195},
Peter M. Garnavich$^2$, Ron L. Gilliland\footnote{Space Telescope Science Institute, 3700 San Martin Drive, 
Baltimore, MD 21218}, Craig J. Hogan$^4$,
Saurabh Jha$^2$, Robert P. Kirshner$^2$,
B. Leibundgut\footnote{European Southern Observatory, Karl-Schwarzschild-Strasse 
2,  Garching, Germany},
M. M. Phillips\footnote{Cerro Tololo Inter-American Observatory,
Casilla 603, La Serena, Chile.  NOAO is operated by the
Association of Universities for Research in Astronomy (AURA) under
cooperative agreement with the National Science Foundation.},
David Reiss$^{4}$, Brian P. Schmidt\footnote{Mount Stromlo and Siding Spring Observatories,
Private Bag, Weston Creek P.O. 2611,  Australia}
\footnote{Visiting astronomer, Cerro Tololo Inter-American
Observatory, National Optical Astronomy Observatories, operated by the
Association of Universities for Research in Astronomy (AURA) under
cooperative agreement with the National Science Foundation.},
 Robert A. Schommer$^7$,
R. Chris Smith$^{7\,}$\footnote{University of Michigan, Department of Astronomy, 834 
Dennison, Ann Arbor, MI 48109},
J. Spyromilio$^{6}$,
Christopher Stubbs$^{4}$,
Nicholas B. Suntzeff$^7$,
John Tonry\footnote{Institute for Astronomy, University of Hawaii, 2680
Woodlawn Dr., Honolulu, HI 96822}

\begin{abstract}
We present spectral and photometric observations of 10 type Ia
supernovae (SNe Ia) in the redshift range 0.16 $\leq z \leq$ 0.62.
The luminosity distances of these objects are determined by methods
that employ relations between SN Ia luminosity and light curve shape.
Combined with previous data from our High-Z Supernova Search Team (Garnavich et al. 1998; Schmidt et
al. 1998) and Riess et al. (1998a), this expanded set of 16 high-redshift supernovae and a set of
34 nearby supernovae are used
to place constraints on the following cosmological parameters: the
 Hubble constant ($H_0$), the mass density ($\Omega_M$), the
cosmological constant (i.e., the vacuum energy density,
$\Omega_\Lambda$), the deceleration parameter ($q_0$), and the dynamical age of the Universe ($t_0$).
The distances of the high-redshift SNe Ia are, on average, 10\% to 15\% farther than expected in a low mass density ($\Omega_M=0.2$) Universe without a cosmological constant.
Different light curve fitting methods, SN Ia subsamples, and prior
constraints unanimously favor
eternally expanding models with
positive cosmological constant (i.e., $\Omega_\Lambda >
0$) and a current acceleration of the expansion (i.e., $q_0
< 0$).
 With no prior constraint on mass density other than  $\Omega_M \geq
0$, the spectroscopically confirmed SNe Ia are statistically
consistent with $q_0 <0$ at the 2.8$\sigma$ and 3.9$\sigma$ confidence
levels, and with $\Omega_\Lambda >0$ at the 3.0$\sigma$ and 4.0$\sigma$
confidence levels, for two different fitting methods respectively. 
Fixing a ``minimal'' mass density, $\Omega_M=0.2$, results in the weakest detection,
$\Omega_\Lambda>0$ at the  3.0$\sigma$ confidence level from one of the
two methods.  For a flat-Universe prior ($\Omega_M+\Omega_\Lambda=1$), the
spectroscopically confirmed SNe Ia require
$\Omega_\Lambda >0$ at  7$\sigma$ and 9$\sigma$ formal
significance for the two different
fitting methods.  A Universe closed by ordinary matter (i.e., $\Omega_M=1$) is formally ruled out at the 7$\sigma$ to 8$\sigma$ confidence level for the two different fitting methods.  We estimate the dynamical age of the Universe to be 14.2 $\pm 1.5$
Gyr including systematic uncertainties in the current Cepheid distance scale.
 We estimate 
the likely effect of several sources of systematic error,
including progenitor  and  metallicity evolution,
extinction, sample selection bias, local perturbations
in the expansion rate, gravitational lensing, and 
sample contamination.  Presently, none of these effects reconciles the data with $\Omega_\Lambda=0$ and $q_0 \geq 0$.
\end{abstract}
subject headings:  supernovae:general ; cosmology:observations

\section{Introduction} 
   This paper reports observations of 10 new high-redshift type Ia
supernovae (SNe Ia) and
the values of the cosmological parameters derived from them.   Together with the 
four high-redshift supernovae previously reported by our High-Z Supernova Search
Team (Schmidt et al. 1998; Garnavich et al. 1998) and two others
(Riess et al. 1998a), the sample of 16 is
now large enough to yield interesting cosmological results of
high statistical significance. Confidence in these results depends
not on increasing the sample size but on improving our
understanding of systematic uncertainties.

    The time evolution of the cosmic scale factor depends on the 
composition of mass-energy in the Universe.  While the Universe is known to contain
a significant amount of ordinary matter, $\Omega_M$, which decelerates
 the expansion, 
its dynamics may also be significantly affected by more 
exotic forms of energy.  Pre-eminent among these is a possible
energy of the vacuum ($\Omega_\Lambda$), Einstein's ``cosmological constant,'' whose negative pressure 
 would do work to accelerate the expansion (Carroll, Press, \& Turner 1992;
Schmidt et al. 1998).  Measurements of 
the redshift and apparent brightness of SN Ia of known intrinsic brightness can constrain these cosmological parameters.

\subsection{The High-Z Program}

 Measurement of the elusive cosmic parameters $\Omega_M$ and 
$\Omega_\Lambda$ through the redshift-distance relation depends on comparing the apparent magnitudes of low-redshift 
SNe Ia with those of their high-redshift cousins.  This requires great
care to assure uniform treatment of both the nearby and distant
samples.

  The High-Z Supernova Search Team has embarked on a program to measure supernovae
at high redshift and to develop the comprehensive understanding of their
properties required for their reliable use in cosmological work.  
Our
team pioneered the use of supernova light curve shapes to 
reduce the scatter about the Hubble line from $\sigma$ $\approx$ 0.4 mag to 
$\sigma$ $\approx$ 0.15 mag (Hamuy et al. 1996a,b, 1995; Riess, Press \&
Kirshner 1995, 1996a).  This dramatic improvement in the
precision of SNe Ia as distance indicators increases the power of
statistical inference for each object by an order of magnitude and
sharply reduces their
susceptibility to selection bias.  Our team has 
also pioneered methods for using multi-color observations to estimate 
the reddening to each individual supernova, near and far, with the aim of 
minimizing the confusion between effects of cosmology and dust (Riess, Press, \& Kirshner
1996a; Phillips et al. 1998).  Because the remaining scatter
 about the Hubble line is so small, the discussion of the Hubble constant 
from low-redshift SNe Ia has already passed into a discussion of the
best use of Cepheid distances to galaxies that have hosted SNe Ia (Saha et al. 
1997; Kochanek 1997; Madore \& Freedman 1998; Riess, Press, \& Kirshner 1996a; Hamuy et al. 1996a; 
Branch 1998).  As the use of SNe Ia for
measuring $\Omega_M$ and $\Omega_\Lambda$ progresses from its infancy into 
childhood, we can
expect a similar shift in the discussion from results limited principally
by statistical errors to those limited by our depth of
understanding of SNe Ia. 

Published high-redshift SN Ia data is a small fraction of 
the data in hand both for our team and for the Supernova Cosmology Project
(Perlmutter et al. 1995, 1997, 1998).  Now 
is an opportune time to spell out 
details of the analysis, since further increasing the sample size without 
scrupulous attention to photometric calibration, uniform treatment of nearby and
distant samples, and an effective way to deal with reddening will not be 
profitable. 
Besides presenting results for four high-z supernovae, we have published details 
of our photometric system (Schmidt et al. 1998) and stated precisely
how we used ground-based photometry to calibrate our {\it Hubble Space
Telescope (HST)} light curves
(Garnavich et al. 1998).  
In this paper, we spell out details of newly-observed 
light curves for 10 objects, explain the recalibration of the relation 
of light curve shape and luminosity for a large low-redshift sample, 
and combine all the data from our team's work to constrain cosmological parameters.
  We also evaluate how systematic effects could 
alter the conclusions.  While some comparison with the stated results 
of the Supernova Cosmology Project (Perlmutter et al. 1995, 1997,
1998) is possible, an informed combination 
of the data will have to await a similarly detailed description of their
measurements.  

\subsection{A Brief History of Supernova Cosmology}

While this paper emphasizes new data and constraints for cosmology, 
a brief summary of the subject may help readers connect work on supernovae
with other approaches to measuring cosmological parameters.

Empirical evidence for SNe I presented by Kowal (1968) showed that these
events had
a well-defined Hubble diagram whose intercept could provide a good 
measurement of the Hubble constant. Subsequent evidence showed that the
original spectroscopic class of Type I should be split (Doggett
\& Branch 1985; Uomoto \& Kirshner 1985; Wheeler \& Levreault 1985;
Wheeler \& Harkness 1986; Porter \& Filippenko 1987).  The remainder
of the original group, now called Type Ia, had peak brightness dispersions
of 0.4 mag to 0.6 mag (Tammann \& Leibundgut 1990; Branch \& Miller
1993; Miller \& Branch 1990; Della Valle \& Panagia 1992; Rood 1994;
Sandage \& Tammann 1993; Sandage et al. 1994).  Theoretical models suggested that these
``standard candles'' arose from the thermonuclear
explosion of a carbon-oxygen white dwarf that had grown to the
Chandrasekhar mass (Hoyle \& Fowler 1960; Arnett 1969; Colgate \&
McKee 1969).  Because SNe Ia are so luminous ($M_B \approx -19.5$ mag),
Colgate (1979) suggested that observations of SNe Ia
at $z \approx 1$ with the forthcoming Space Telescope could measure the
deceleration parameter, $q_0$.

    From a methodical CCD-based supernova search
that spaced observations across a lunation and employed prescient use of
image-subtraction techniques to reveal new objects,  Hansen,
N\o rgaard-Nielsen, \& Jorgensen (1987) detected SN 1988U, a SN Ia at
$z=0.31$ (N\o rgaard-Nielsen et al. 1989).    At this redshift
and distance precision ($\sigma \approx 0.4 $ to 0.6 mag),
$\sim 100$ SNe Ia would have been needed to distinguish between an open and
closed Universe.  Since the Danish group had already spent two years to find one object,
it was clear that larger detectors and faster telescopes needed to be
applied to this problem.

  Evidence of systematic problems also lurked in supernova photometry
so that merely increasing the sample would not be adequate.  
Attempts to correct supernova magnitudes for reddening by 
dust (Branch \& Tammann 1992) based on the plausible (but incorrect) assumption 
that all SNe Ia had the same
intrinsic color had the unfortunate effect of increasing
the scatter about the Hubble line or alternately attributing bizarre
properties to the dust absorbing SN Ia light in other galaxies.  In addition,
well-observed supernovae such as SN 1986G (Phillips et al. 1987;
Cristiani et al. 1992), SN 1991T (Filippenko et
al. 1992a; Phillips et al. 1992; Ruiz-Lapuente et al. 1992), and SN 1991bg (Filippenko et al. 1992b; 
Leibundgut et al. 1993; Turatto et al. 1996)
indicated that large and real inhomogeneity was buried in the scatter about the Hubble line. 

   Deeper understanding of low-redshift supernovae greatly improved their cosmological 
utility.    Phillips (1993)
reported that the observed peak luminosity of SNe Ia varied by a factor of
3.  But he also showed that the decrease in
$B$ brightness in the 15 days after peak ($\Delta m_{15}(B)$) was a good predictor of the SN Ia 
luminosity, with slowly declining
supernovae more luminous than those that fade rapidly.  

     A more extensive database of carefully and uniformly observed SNe
Ia was needed to refine the understanding of SN Ia light curves. The
Cal\'{a}n/Tololo survey (Hamuy et al. 1993a) made a systematic photographic
search for supernovae between cycles of the full moon. This search
was extensive enough to guarantee the need for scheduled follow-up
observations, which were supplemented by the cooperation of visiting observers, to 
collect well-sampled
light curves.  Analysis of the Cal\'{a}n/Tololo results generated a
broad understanding of SNe Ia and demonstrated their remarkable
distance precision (after template fitting) of $\sigma \approx 0.15$ mag (Hamuy et al. 1995,
1996a,b,c,d).  A parallel effort employed data from the Cal\'{a}n/Tololo
survey and from the Harvard-Smithsonian Center for Astrophysics (CfA) to develop detailed empirical
models of SN Ia light curves (Riess, Press, \& Kirshner 1995; Riess 1996).  This work was
extended into the Multi-Color Light Curve Shape (MLCS) method which employs
up to 4 colors of SN Ia photometry to yield excellent distance
precision ($\approx 0.15$ mag) and a statistically valid estimate of the uncertainty for
each object with a measurement of the reddening by
dust for each event (Riess, Press, \& Kirshner 1996a;
Appendix of this paper).  This work has also placed useful
constraints on the nature of dust in other galaxies (Riess, Press, \& Kirshner 1996b).  

   The complete sample of nearby SNe Ia light curves from the
Cal\'{a}n/Tololo and CfA samples provides a solid foundation from which to
extend the redshift-distance relation to explore cosmological
parameters.  The low-redshift sample used here has 34 SNe Ia with $z < 0.15$.

Since the high-redshift observations reported here consumed 
large amounts of observing time at the world's finest telescopes, we have
a strong incentive to find efficient ways to use the minimum set of
observations to derive the distance to each supernova.  A recent exploration
of this by Riess et al. (1998a) is the ``Snapshot" method which uses only a
single spectrum and a single set of photometric measurements to infer the luminosity
distance to a SN Ia with $\sim$ 10\% precision.
  In this paper, we employ the snapshot method for six SNe Ia with
sparse data, but a shrewdly designed program that was intended to use
the snapshot approach could be even more effective in extracting useful
results from slim slices of observing time.

   Application of large-format CCDs and sophisticated image analysis
techniques by the Supernova Cosmology Project (Perlmutter et al. 1995)
led to the discovery of SN 1992bi ($z=0.46$) followed by 6 more SNe Ia
at $z \approx 0.4$ (Perlmutter et al. 1997).  Employing a correction for
the luminosity/light-curve shape relation (but none for host galaxy
extinction), comparison of these SNe Ia to the Cal\'{a}n/Tololo sample gave an 
initial indication of
``low'' $\Omega_\Lambda$ and ``high'' $\Omega_M$:
$\Omega_\Lambda=0.06^{+0.28}_{-0.34}$ for a flat Universe and
$\Omega_M=0.88^{+0.69}_{-0.60}$ for a Universe without a cosmological
constant ($\Omega_\Lambda \equiv 0$).  The addition of one very 
high-redshift ($z=0.83$) SN Ia
observed with {\it HST} had a significant effect on the results:
$\Omega_\Lambda=0.4\pm0.2$ for a flat Universe, and
$\Omega_M=0.2\pm0.4$ for a Universe with $\Omega_\Lambda \equiv 0$.  
(Perlmutter et al. 1998). This illustrates how young and volatile the subject is 
at present. 

\subsection{This Paper}

  Our own High-Z Supernova Search Team has been assiduously discovering high-redshift
supernovae, obtaining their spectra, and measuring their light curves
since 1995 (Schmidt et al. 1998).  The goal is to provide
an independent set of measurements that uses our own techniques
and compares our data at high and low redshifts to constrain the
cosmological parameters.   Early results from 4 SNe Ia (3 observed
with {\it HST}) hinted at a non-negligible cosmological constant and ``low''
$\Omega_M$, but were limited by statistical errors: $\Omega_\Lambda=0.65 \pm 0.3$
for a flat Universe, $\Omega_M=-0.1\pm0.5$ when $\Omega_\Lambda \equiv
0$ (Garnavich et al. 1998).  Our aim in this paper is
to move the discussion forward by increasing the data set from four
high-redshift SNe to 16, to spell out exactly how we have made the measurement, and to
consider various possible systematic effects.  

In \S 2 we describe the observations of the SNe Ia including
their discovery, spectral identification, photometric calibration, and
light curves.  We determine the luminosity
distances (including $K$-corrections) via two methods, MLCS and a
template fitting method ($\Delta m_{15}(B)$), as explained in \S 3.  Statistical inference
of the cosmological parameters including $H_0$, $\Omega_M$,
$\Omega_\Lambda$, $q_0$, $t_0$, and the fate of the
Universe is contained in \S 4.  Section 5 presents a quantitative
discussion of systematic uncertainties which could affect our results:
evolution, absorption, selection bias, a local void, weak lensing, and
sample contamination.  Our conclusions are
summarized in \S 6.

\section{Observations}
\subsection{Discovery}

   We have designed a search program to find supernovae in the
redshift range $0.3<z<0.6$ with the purpose of measuring
luminosity distances to constrain cosmological parameters
(Schmidt et al. 1998). Distances with the highest precision are measured from SNe Ia
observed before maximum brightness and in the redshift range of $0.35 < z < 
0.55$, where 
our set of custom passbands measures the supernova light emitted in rest-frame $B$ and $V$.  
By imaging fields near 
the end of a
dark run, and then again at the beginning of the next
dark run, we
ensure that the newly discovered supernovae are young
(N\o rgaard-Nielsen et al. 1989; Hamuy et al. 1993a;
Perlmutter et al. 1995).  Observing a large area
and achieving a limiting magnitude of 
$m_R\approx23$~mag yields many SN Ia candidates in the
desired redshift range
(Schmidt et al. 1998).   By obtaining spectra of these candidates with
4-m to 10-m telescopes, we can identify the SNe Ia and confirm their youth
using the spectral feature aging technique of Riess et al. (1997).

The 10 new SNe Ia presented in this paper (SN 1995ao, SN
1995ap, SN 1996E, SN 1996H, SN 1996I, SN 1996J, SN 1996K, SN 1996R, SN
1996T, and SN 1996U) were discovered 
using the CTIO Blanco 4 m telescope with the facility prime-focus CCD camera as part of a 3-night
program in 1995 Oct-Nov and a 6-night 
program in 1996 Feb-Mar. This instrument
has a pixel scale of 0.43$^{\prime\prime}$, and the TEK 2048x2048 pixel CCD frame covers 0.06 square 
degrees.  In each of the search programs, multiple images were
combined after removing cosmic rays, differenced with ``template'' images,
and searched for new objects using the prescription of \markcite{S98}Schmidt et al. (1998).
The data on 1995 Oct 27 and 1995 Nov 17 were gathered under mediocre conditions with most images having seeing worse
than $1.5^{\prime\prime}$.   The resulting differenced images were sufficient to find new objects brighter
than $m_R=22.5$ mag. The data acquired in 1996 had better image quality
($\sim 1.5^{\prime\prime}$), and the
differenced images were sufficient to uncover new objects brighter than $m_R=23$~mag.

In total, 19 objects were identified as possible supernovae ---  2 new objects were detected on both 
1995 November 17 and 1995 November 29, 5 new objects
on 1996 Feb 14-15, 2 on 1996 Feb 20-21, and 8 on 1996 Mar 15-16
(Kirshner et al. 1995; Garnavich et al. 1996a,b).

\subsection{Data}

Spectra of the supernova candidates were obtained to
classify the SNe and obtain redshifts of their host galaxies. For this purpose, the Keck telescope,
Multiple-Mirror Telescope (MMT), and the European Southern Observatory
3.6-m (ESO~3.6-m)
were utilized following the Fall 1995 and Spring 1996 search campaigns.  Some galaxy redshifts were obtained with the Keck telescope in the Spring of 1998.

The Keck spectra were taken with the Low Resolution Imaging Spectrograph
(LRIS; Oke et al. 1995), providing a resolution of 6 \AA\ full width
at half maximum (FWHM). Exposure times were
between 3$\times$900 seconds and 5$\times$900 seconds, depending on the
candidate brightness. 

The MMT spectra were obtained
with the Blue Channel spectrograph and 500 lines/mm grating giving
a resolution of 3.5 \AA\ FWHM. Exposure times were 1200 seconds and repeated
five to seven times. The MMT targets were placed
on the slit using an offset from a nearby bright star. 

The ESO~3.6-m data were collected
with the ESO Faint Object Spectrograph Camera (EFOSC1) at
a nominal resolution of 18 \AA\ FWHM.  Single 2700 second exposures
were made of each target.

Using standard reduction packages in IRAF, the CCD images were bias
subtracted and divided by a flat-field
frame created from a continuum lamp exposure. Multiple images
of the same object were shifted where necessary and combined
using a median algorithm to remove cosmic ray events. For
single exposures, cosmic rays were removed by hand using the IRAF/IMEDIT
routine. Sky emission lines were problematic, especially longward
of 8000 \AA\ .  The spectra were averaged perpendicular to the dispersion
direction, and that average was subtracted from each line along the
dispersion. However, residual
noise from the sky lines remains. The one-dimensional
spectra were then extracted using the IRAF/APSUM routine and
wavelength calibrated either
 from a comparison lamp exposure or
the sky emission lines. The flux was calibrated using observations of standard stars
and the IRAF/ONEDSTDS database.

The candidates were classified from visual inspection of
their spectra and comparison with the spectra of well-observed
supernovae (see \S 5.7). In all, 10 of the candidates were SNe Ia,
1 was a SN II, and 2 were active galactic nuclei or SNe II
(\markcite{K95}Kirshner et al. 1995; \markcite{G96a}\markcite{G96b}Garnavich et al. 1996a,b). The remaining 6 candidates were observed, but the spectra did not have sufficient signal to
allow an unambiguous classification.  The identification spectra for
the 10 new SNe Ia are summarized in Table 1 and shown in Figure 1.  In
addition we include the spectral data for 3 previously analyzed SNe:
SN 1997ce, SN 1997cj, and SN 1997ck (Garnavich et al. 1998).  The
spectral data for SN 1995K are given by Schmidt et al. (1998).  The spectrum
 of SN 1997ck shows only a [O II] emission line at 7328.9 \AA\ in four separate
exposures (Garnavich et al. 1998).  The equivalent $R$ band magnitude of the exposure was 26.5 which is more than 1.5 magnitudes dimmer than the supernova would have been in $R$, suggesting that the SN was not in the slit when the host galaxy was observed.

Most of the host galaxies showed emission
lines of [O II], [O III], or H$\alpha$ in the spectrum, and the redshift
was easily measured for these. For the remainder, the redshift was
found by matching the broad features in the high-redshift supernovae to
those in local supernova spectra.  The intrinsic dispersion in the expansion velocities of SNe Ia (Branch et al. 1988; Branch \& van den Bergh 1993)
limits the precision of this method to $1\sigma \approx 2500$ km
s$^{-1}$ independent of the signal-to-noise ratio of the SN spectrum.
The method used to determine the redshift for each SN is given in
Table 1.

  Following the discovery and identification of the SNe Ia, photometry of
these objects was obtained from 
observatories scheduled around the world.  The SNe were primarily observed through custom
passbands designed to match the
wavelength range closest to rest-frame Johnson $B$ and $V$ passbands.
Our ``$B45$,'' ``$V45$,'' ``$B35$,'' and ``$V35$'' filters are specifically
designed to match Johnson $B$ and $V$ redshifted by $z=0.45$ and
$z=0.35$, respectively.  The characteristics of these filters are
described by Schmidt et al. (1998).  A few observations were obtained through standard bandpasses as noted
in Table 2 where we list the
photometric observations for each SN Ia.

 Photometry of local standard stars in the supernova fields in the
 $B35,V35,B45,V45$ (or ``supernova")
photometric system were derived from
data taken on three photometric nights.  The method has been
described in Schmidt et al. (1998) but we summarize it here.
 The supernova photometric system has
been defined by integrating the fluxes of spectrophotometric standards from Hamuy et al. (1994) through the supernova bandpass response functions (based on the filter
transmissions and a typical CCD quantum efficiency function) and solving for the
photometric coefficients which would yield zero color for these stars and monochromatic magnitudes of 0.03 for Vega. 

 This {\it theoretically} defined
photometric system also provides transformations between the
Johnson/Kron-Cousins system and the supernova system. We use
theoretically derived transformations to convert the known $V,R,$ and
$I$ magnitudes of Landolt (1992) standard fields into $B35,V35,B45,V45$ photometry.

On nights which are photometric, we observe Landolt standard fields with the
$B35,V35,B45,V45$ filters and measure the stars' instrumental magnitudes from apertures
 large enough to collect all the stellar light.  We then derive the
transformation from the supernova system to the instrumental system as
a function of the instrumental magnitudes, supernova system colors,
and observed airmass.  Because our theoretical response functions are
very similar to the instrumental response functions, our measured
color coefficients were small, typically less than 0.02 mag per
mag of $B45-V45$ or $B35-V35$.  These long wavelength filters also
reduced the effect of atmospheric extinction (compared to $B$ and $V$).  Typical extinction
coefficients were 0.11, 0.09, 0.07, and 0.06 mag per airmass for $B35$,
$B45$, $V35$, and $V45$, respectively.

Isolated stars on each supernova frame were selected as local
standards.  The magnitudes of the local standards were determined from
the transformation of
their instrumental magnitudes, measured from similarly large apertures.
The final transformed magnitudes of these local standards, averaged
over three photometric nights, is given in Table 3.  The locations of
the local standards and the SNe are shown in Figure 2.
The uncertainties in the local standards' magnitudes are the quadrature
sum of the uncertainty (dispersion) of the instrumental transformations
(typically 0.02 mag) and the individual uncertainties from photon
(Poisson) statistics.  The dispersion in the instrumental
transformation quantifies the errors due
 to imperfect flat-fielding, small changes in the
atmospheric transparency, incomplete empirical modeling of the
response function, and seeing variations.  This uncertainty is valid
for any single observation of the local standards.

  To measure the brightness of the supernovae free from host galaxy
contamination, we obtained deep images of the hosts a year
after, or months before, the discovery of the SNe.  These images were
used to subtract digitally a host's light from
the supernova's light, leaving only the stellar point spread function (PSF).  The algorithms employed
to match the resolution, intensity, and coordinate frames of images
prior to subtraction are described in Schmidt et al. (1998).  The
brightness of the
SNe in these uncrowded fields was then measured relative to the calibrated local standard stars in the field by
fitting a model of a PSF to the stars and supernova using the DoPHOT algorithm
(Schmidt et al. 1998; Mateo \& Schechter 1989; Schechter, Mateo, \& Saha
1993).  

  Systematic and statistical
components of error were evaluated by measuring the brightness of artificial
stars added to the subtracted frames.  These artificial stars had the same brightness and background as
the measured SNe (Schmidt et al. 1998).
The ``systematic'' error was measured from the difference in the
mean magnitude of the artificial stars {\it before} and {\it after} the image
processing (i.e., alignment, scaling, ``blurring,'' and subtracting).
The systematic errors were always $<$ 0.1 mag and were of
either sign.  Any significant
systematic error is likely the result of a mismatch in the global
properties of the template image and SN image based on only examining a local region of the two images.   A correction based
on the systematic error determined from the artificial stars was
applied to the measured SN magnitude to yield an
unbiased estimate of the SN magnitude.  The dispersion of the recovered
artificial magnitudes about their mean was assigned to the statistical
uncertainty of the SN magnitude.  

The supernova PSF magnitudes were transformed to the $B35,V35,B45,V45$
system using the local standard magnitudes and the color coefficients derived from
observations of the Landolt standards.  The final SN light
curves are the average of the results derived from 5 or 6 local
standards, weighted by the uncertainty of each local standard star.  The light
curves are listed in Table 4 and displayed in Figure 3.
The SN magnitude errors are derived from the artificial star
measurements as described above.

  The small color and atmospheric
extinction coefficients give us confidence that the supernova
photometry 
accurately transformed to the
$B35,V35,B45,V45$ system.  However, it is well known that a nonstellar
flux distribution can produce substantial systematic errors in supernova
photometry (Menzies 1989). We have anticipated this
problem by using identical filter sets at the various observatories and
by defining our photometric system with actual instrumental response
functions. To measure the size of this effect on our SN photometry,
 we have calculated the systematic error incurred from the differences
in the instrumental response functions of different observatories we
employed.  Spectrophotometric calculations from SN Ia spectra
using various instrumental response functions show that the expected
differences are less than 0.01 mag and can safely be ignored.

\section{Analysis}

\subsection{$K$-corrections}

    A strong empirical understanding of SN Ia light curves has
been garnered from intensive monitoring of SNe Ia at $z \leq 0.1$
through $B$ and $V$ passbands (Hamuy et al. 1996b; Riess 1996; Riess et
al. 1998b; Ford et al. 1993;
 Branch 1998, and
references therein).
 We use this
understanding to compare the light curves of the high-redshift and low-redshift samples at the same rest wavelength.  By a judicious choice
of filters, we minimize the differences between $B$ and $V$
rest-frame light observed for distant SNe and their nearby
counterparts.  Nevertheless, the range of redshifts involved makes it
difficult to eliminate all such differences.  We therefore employ
``$K$-corrections'' to convert the observed magnitudes to rest-frame $B$
and $V$ (Oke \& Sandage 1968; Hamuy et al. 1993b; Kim, Goobar, \& Perlmutter 1996; Schmidt et al. 1998).

  The cross-band $K$-correction for SNe Ia has been described as a function of the observed and rest-frame filter transmissions, the
redshift of the supernova, and the age of the supernova (see equation
1 of Kim, Goobar, \& Perlmutter 1996).  Such a $K$-correction assumes that the spectral energy distribution of all SNe Ia of a given
age is homogeneous,  yet it has been shown (Pskovskii 1984; Phillips et al. 1987, 1993; Leibundgut
et al. 1993;  Nugent et al.
1995; Riess, Press, \& Kirshner 1996a; Phillips et al. 1998; Lira 1995;  
Appendix of this paper) that at a given age, the colors of SNe Ia
exhibit real variation related to the absolute magnitude of the supernova.  

   A variation in SN Ia color, at a fixed phase, could have dire consequences for
determining accurate $K$-corrections.  An appropriate $K$-correction 
quantifies the {\it difference} between the supernova light which falls
into a standard passband (e.g., $B$) at $z=0$ and that which falls into
the filters we employ to observe a redshifted SN Ia.  Differences
in SN Ia color, at a fixed phase, would alter the appropriate
$K$-correction.  We need to know the color of each supernova to
determine its $K$-correction precisely.
Differences in SN Ia color can arise from interstellar extinction or intrinsic properties of the supernova such
as a variation in photospheric temperature (Nugent et al. 1995).  

Nugent et al. (1998) have shown that, to within 0.01 mag, both the
effects of extinction and intrinsic variations on the SN Ia spectral energy
distribution near rest-frame $B$ and $V$ and hence on the $K$-correction can be reproduced by application of a Galactic reddening
law (Cardelli, Clayton, \& Mathis 1989) to the spectra.  The difference in color, at a given age, between an individual SN Ia and a
fiducial SN Ia is quantified by a color excess, $E_{B-V}$, and determines
the effects of either extinction or intrinsic variation on the spectra
and observed colors of the SNe.  For most epochs, filter combinations,
and redshifts, the variation of the $K$-correction with the observed
variations of color excess is only 0.01 to 0.05 mag.  For redshifts
which poorly match the rest-frame wavelengths to the observed
wavelengths, the custom $K$-correction for very red or very blue SNe Ia can differ
from the standard $K$-correction by 0.1 to 0.2 mag.  

  This prescription requires the age and observed color for each
observation to be known before its $K$-correction
can be calculated.  The age is best determined from fitting the light curve's
time of maximum.  Yet we must use the $K$-correction to determine the
time of maximum and the true color of each epoch.  This conundrum can
be solved by iteratively converging
to a solution by repeated cycles of $K$-correcting and empirical
fitting of the light
curves.  Table 4 lists the final cross-band $K$-corrections we used
to convert the observations to the rest-frame passbands.

\subsection{Light Curve Fitting}

   As described in \S 1, empirical models for SNe Ia light curves which
employ the observed correlation between light curve shape,  luminosity, and color have led to significant improvements
in the precision of distance estimates derived from SNe Ia (Hamuy et al. 1995, 1996a; Riess, Press, \& Kirshner 1995, 1996a).  Here we employ
the MLCS method prescribed by Riess, Press, \& Kirshner 1996a as reanalyzed in the Appendix and the template fitting method
of Hamuy et al. (1995, 1996d) to fit the light curves in Table 4.

   The growing sample of well-observed SN Ia light curves
(Hamuy et al. 1996b; Riess 1996; Riess et al. 1998b; Ford et al. 1993)
justifies refinements in the MLCS method that are described in the
Appendix.  These include a new derivation of
the relation between light curve shape, luminosity, and color from
SNe Ia in the Hubble flow using redshift as the distance indicator.
In addition, this empirical description has been extended to a second order (i.e., quadratic)
relation between SN Ia luminosity and light curve shape.   A more
realistic {\it a priori} probability distribution for extinction has
been utilized from the calculations of Hatano, Branch, \& Deaton (1997).  Further, we
now quantify the residual correlations between 
observations of dissimilar time, passband, or both.   The empirical
model for a SN Ia light
and color curve is still described by four
parameters: a date of maximum ($t$), a luminosity difference ($\Delta$), an
apparent distance ($\mu_B$),
and an extinction ($A_B$). Due to the redshifts of the SN
host galaxies we first correct the
supernova light curves for Galactic extinction (Burstein \& Heiles
1982), then determine host galaxy extinction. 

 To treat the high and low
redshift SNe Ia consistently, we restricted the MLCS fits to the
nearby SNe Ia observations in $B$ and $V$ within 40 days after
maximum brightness in the restframe.  This is the age by which all high-redshift light curve
observations ended.  Because of this restriction, we also limited our 
consideration of nearby SNe Ia to those with light curves which began
no later than $\sim$ 5 days after $B$ maximum.
Although more precise distance estimates could be
obtained for the nearby sample by including later data and additional
colors, the nearby sample is large enough to determine the
nearby expansion rate to sufficient precision.  
The parameters of the MLCS fits to 27 SNe Ia in the
nearby Hubble flow (0.01 $<$ $z$ $<$ 0.13; Hamuy
et al. 1996b; Riess et al. 1998b) are given in Table 10.  

   In Table 5 we list the parameters of the MLCS fits to six SN
Ia light curves presented here (SNe 1996E, 1996H, 1996I, 1996J, 1996K, 1996U) and for three SNe Ia from our previous
work (SNe 1995K, 1997ce, 1997cj; Garnavich et al. 1998; Schmidt et al.
1998).   We have placed all MLCS distances on the Cepheid distance
scale using Cepheid distances to galaxies hosting photoelectrically
observed SNe Ia: SN 1981B, SN 1990N,
and SN 1972E (Saha et al. 1994, 1997; Riess, Press, \& Kirshner 1996a).  However, conclusions
about the values of the cosmological parameters
$\Omega_M$, $\Omega_\Lambda$, and $q_0$ are {\it independent} of the
distance scale. 

  An additional supernova, SN 1997ck, was studied by Garnavich et al.
(1998) in a galaxy with $z$=0.97.  Its rest-frame $B$ light curve was measured
with the {\it HST} (see Figure 3).  Although this object lacks
a spectroscopic classification and useful color
information, its light curve shape and peak luminosity are consistent
with those of a typical SN Ia.  Due to the uncertainty in this object's
extinction and classification, we will analyze the SNe Ia distances both with
and without this most distant object.

We have also determined the distances to the same 27 nearby SNe Ia and
the ten well-observed high-redshift events using a template fitting approach 
(Hamuy et al. 1995, 1996d).  The maximum light magnitudes and the initial 
decline rate parameter $\Delta m_{15}(B)$ for a given SN Ia are derived by comparing the 
goodness-of-fits of the photometric data to a set of six template SN Ia light 
curves selected to cover the full range of observed decline rates.  The 
intrinsic luminosity of the SN is then corrected to a standard value of
the decline rate ($\Delta m_{15}(B)$=1.1) using a linear relation
between $\Delta m_{15}(B)$
and the luminosities for a set of SNe Ia in the Hubble flow (Phillips
et al. 1998).  An extinction correction has been applied to these distances 
based on the measured color excess at maximum light using the relation between $\Delta m_{15}(B)$ and the unreddened SN Ia color at maximum light (Phillips
et al. 1998). These extinction measurements employ the same Bayesian
filter (in the Appendix) used for the 
MLCS fits.  The final distance moduli are also 
on the Cepheid distance scale as described by Hamuy et al. (1996a) and Phillips
et al. (1998).  Parameters of these fits to the nearby and high-redshift SNe Ia
are provided in Table 10 and Table 6, respectively.

   For both the MLCS and template fitting methods, the fit to the
data determines the light curve parameters and 
their uncertainties.  The ``goodness'' of the
fits was within the expected statistical range with the exception of SN 1996J. This supernova is at a measured redshift of $z$=0.30, but
some of the observations were obtained through a set of filters
optimized for $z$=0.45.  The uncertainty from this mismatch and the 
additional uncertainty from separate calibrations of the local standards' magnitudes
in two sets of filters may be the source of the poor result for this
object.

 Four remaining SNe Ia presented here (SNe 1995ao, 1995ap, 1996T, and 1996R) are too
sparsely sampled to provide meaningful light curves fit by either of
the light curve fitting methods.  However, Riess et al. (1998a) describe a technique to measure the
distance to sparsely observed SNe which lack well-sampled light curves.  This ``Snapshot''
method measures the age and the luminosity/light-curve shape parameter
from a SN Ia spectrum using techniques from Riess et al. (1997) and
Nugent et al. (1995).  An additional photometric epoch in 2 passbands
(with host galaxy templates if needed) provides enough
information to determine the extinction-free distance.  For the four
sparsely observed SNe Ia in our sample, we have measured the SN parameters with
this method and list them in Table 7.  This sample of sparsely observed, high-redshift SNe Ia is augmented by distances for SN 1997I ($z=0.17$) and SN 1997ap ($z=0.83$) from Riess et al. (1998a).

   For all SN Ia distance measurements, the dominant source of
 statistical uncertainty
 is the extinction measurement.  The precision of our determination of the true extinction is
improved using our prior understanding of its magnitude and direction (Riess, Press, \& Kirshner 1996a, Appendix).
  
\section {Cosmological Implications of SNe Ia}

  \subsection {Cosmological Parameters}

      Distance estimates from SN Ia light curves are derived from the
luminosity distance, \bq D_{L} = \left(\frac{{\cal L}}{4 \pi {\cal
F}}\right)^{\frac{1}{2}}, \eq where ${\cal L}$ and ${\cal
F}$ are the SN's intrinsic
luminosity and observed flux, respectively.   In Friedmann-Robertson-Walker cosmologies, the 
luminosity distance at a given redshift, $z$, is a function of the
cosmological parameters.  Limiting our consideration of these
parameters to the Hubble
constant, $H_0$, the mass density, $\Omega_M$, and the vacuum energy
density (i.e., the cosmological constant), $\Omega_\Lambda$ (but see
Caldwell, Dave, \& Steinhardt 1998; Garnavich et al. 1998 for other
energy densities), the
luminosity distance is
\bq
 D_L= c H_0^{-1}(1+z)\left | \Omega_k \right |^{-1/2}sinn\lbrace\left | \Omega_k \right |^{1/2}
\int_0^zdz[(1+z)^2(1+\Omega_Mz)-z(2+z)\Omega_\Lambda ]^{-1/2}\rbrace,
\eq
where $\Omega_k=1-\Omega_M-\Omega_\Lambda$, and $sinn$ is $\sinh$ for
$\Omega_k \geq 0$ and $\sin$ for $\Omega_k \leq 0$
(Carroll, Press, \& Turner 1992).  For $D_L$ in units of 
Megaparsecs, the predicted distance modulus is \bq \mu_p=5\log D_L +25
.\eq

Using the data described in \S 2 and the fitting methods of
\S 3 we have derived a set of distances, ${\bf \mu_0}$, for SNe with $0.01 \leq z
\leq 0.97$.  The available set of high-redshift SNe includes nine well-observed SNe
Ia, six sparsely observed SNe Ia, and SN
1997ck ($z=0.97$) whose light curve was well observed but lacks
spectroscopic classification and color measurements.
The Hubble diagrams for the nine well-observed SNe Ia plus SN 1997ck, with light curve distances calculated from
the MLCS method and the template approach, are shown in Figures 4 and
5.  The likelihood for the cosmological parameters can be determined
from a $\chi^2$ statistic, where \bq \chi^2(H_0,\Omega_M,\Omega_\Lambda)=\sum_i { (\mu_{p,i}(z_i; H_0, \Omega_m, \Omega_\Lambda)-\mu_{0,i})^2 \over
\sigma_{\mu_{0,i}}^2+\sigma_v^2} \eq and $\sigma_v$ is the dispersion
in galaxy redshift (in units of distance moduli) due to peculiar
velocities. This term also includes the uncertainty in galaxy
redshift.  We have
calculated this $\chi^2$ statistic for a wide range of the parameters
$H_0$, $\Omega_M$, and $\Omega_\Lambda$.  We do not consider the unphysical region of parameter space where
$\Omega_M < 0$; equation (2) describes the effect of massive particles
on the luminosity distance.  There is no reason to expect that the evaluation of
equation (2) for $\Omega_M < 0$ has any correspondence to physical
reality.  We also neglect the region of ($\Omega_M,\Omega_\Lambda$)
parameter space which gives rise to so-called ``bouncing'' or
rebounding Universes which do not monotonically expand from a ``big bang'' and
for which equation (2) is not solvable (see
Figure 6 and 7) (Carroll, Press, \& Turner 1992). 

   Due to the large redshifts of our distant sample and the abundance
of objects in the nearby sample, our analysis is insensitive to $\sigma_v$ within its likely range of 100 km s$^{-1}$ $\leq \sigma_v \leq$
400 km s$^{-1}$ (Marzke et al. 1995; Lin et al. 1996).  For our
analysis we adopt $\sigma_v=200$ km s$^{-1}$.  For high-redshift SNe
Ia whose redshifts were determined from the broad features in the SN
spectrum (see Table 1), we add 2500 km s$^{-1}$ in quadrature to $\sigma_v$.

   Separating the effects of matter density and vacuum
energy density on the observed redshift-distance relation could in
principle be accomplished with
measurements of SNe Ia over a significant range of high redshifts (Goobar \&
Perlmutter 1995).  Because the matter density decreases with time in
an expanding Universe while the vacuum energy density remains
constant, the relative influence of $\Omega_M$
to $\Omega_\Lambda$ on the redshift-distance relation is a function of
redshift. The present
data set has only a modest range of redshifts so we can only constrain specific cosmological models or regions of
($\Omega_M$, $\Omega_\Lambda$)
parameter space to useful precision.

The $\chi^2$ statistic of equation (4) is well suited for determining
the most likely values for the cosmological parameters $H_0$,
$\Omega_M$, and $\Omega_\Lambda$ as well as the confidence intervals
surrounding them.  For constraining regions of parameter space not
bounded by contours of uniform confidence (i.e., constant $\chi^2$),
we need to define the probability density function (PDF) for the
cosmological parameters.  The PDF ($p$) of these parameters given our distance moduli is derived from the PDF of the distance moduli given our data from Bayes' theorem, \bq 
p(H_0,\Omega_m,\Omega_\Lambda | {\bf \mu_0}) = \frac{p({\bf \mu_0} |
H_0, \Omega_m, \Omega_\Lambda) p(H_0, \Omega_m, \Omega_\Lambda)}{p({\bf \mu_0})},
 \eq where ${\bf \mu_0}$ is our set of distance moduli (Lupton 1993).
Since we have no prior constraints on the cosmological parameters
(besides the excluded regions) or on the data, we
take $p(H_0,\Omega_m,\Omega_\Lambda)$ and $p({\bf \mu_0})$ to be constants.  Thus we have for the allowed region of $(H_0,\Omega_m,\Omega_\Lambda)$ 

\bq 
p(H_0,\Omega_m,\Omega_\Lambda | {\bf \mu_0}) \propto 
p({\bf \mu_0} | H_0, \Omega_m, \Omega_\Lambda). 
 \eq

   We assume each distance modulus is independent (aside from
systematic errors discussed in \S 5) and normally
distributed, so the PDF for the set of distance moduli given the
parameters is a product of Gaussians:

\bq 
p({\bf \mu_0} | H_0, \Omega_m, \Omega_\Lambda) =
\prod_i \frac{1}{\sqrt{2 \pi (\sigma_{\mu_{0,i}}^2+\sigma_v^2)}} \exp\left(-\frac{[\mu_{p,i}(z_i; H_0, \Omega_m, \Omega_\Lambda)-\mu_{0,i}]^2}{2
(\sigma_{\mu_{0,i}}^2+\sigma_v^2)}\right).
 \eq  Rewriting the product as a summation of the exponents and
combining with equation (4) we have

\bq
p({\bf \mu_0} | H_0, \Omega_m, \Omega_\Lambda) =
\left(\prod_i \frac{1}{\sqrt{2 \pi (\sigma_{\mu_{0,i}}^2+\sigma_v^2)}} \right)
\exp \left( - \frac{\chi^2}{2} \right).
\eq  The product in front is a constant, so combining with equation
(6) the PDF for the cosmological parameters yields the standard
expression (Lupton 1993)

\bq p(H_0,\Omega_m,\Omega_\Lambda | {\bf \mu_0}) \propto \exp \left( -
\frac{\chi^2}{2} \right). \eq  The normalized PDF comes from dividing this relative PDF by its sum over all possible states,

\bq p(H_0,\Omega_m,\Omega_\Lambda | {\bf \mu_0}) =
\frac{\exp \left( -\frac{\chi^2}{2} \right)}
{  \int_{-\infty}^\infty \, dH_0 \, \int_{-\infty}^\infty d\Omega_\Lambda
\int_0^\infty \exp \left( -\frac{\chi^2}{2} \right) \, d\Omega_M },
\eq neglecting the unphysical regions.  The most likely values for the cosmological parameters and preferred
regions of parameter space are located where equation (4) is minimized or
alternately equation (10) is maximized.  

   The Hubble constants as derived from the MLCS method, 65.2 $\pm1.3$ km
s$^{-1}$ Mpc$^{-1}$, and from the template fitting
approach, 63.8 $\pm1.3$ km s$^{-1}$ Mpc$^{-1}$, are extremely robust
and attest to the consistency of the methods.  These determinations include {\it only} the statistical component of error resulting from the point-to-point variance of the measured Hubble flow and do not include any uncertainty in the absolute magnitude of SN Ia.  
From three photoelectrically observed SNe Ia, SN 1972E, SN 1981B, and SN 1990N (Saha et
al. 1994, 1997), the SN Ia absolute magnitude
was calibrated from observations of Cepheids in the host
galaxies.  The calibration of the SN Ia magnitude from only three
objects adds an additional 5\% uncertainty to the Hubble constant,
independent of the uncertainty in the zeropoint of the distance scale.
The uncertainty in the Cepheid distance scale adds an
uncertainty of $\sim$ 10\% to the derived Hubble constant (Feast \& Walker 1987; Kochanek 1997; Madore \&
Freedman 1998).  A realistic determination of the Hubble constant 
from SNe Ia would give 65 $\pm$ 7 km s$^{-1}$ Mpc$^{-1}$ with the
uncertainty dominated by the systematic uncertainties in the
calibration of the SN Ia absolute magnitude.  These determinations of
the Hubble constant employ the Cepheid distance scale of Madore \&
Freedman (1991) which uses a distance modulus to the Large Magellanic
Cloud (LMC) of 18.50 mag.  Parallax measurements by the {\it
Hipparcos} satellite indicate that the LMC distance could be greater
and hence our inferred Hubble constant smaller by 5\% to 10\% (Reid
1997), though not all agree with the interpretation of these
parallaxes (Madore \& Freedman 1998).  All subsequent
indications in this paper for the cosmological parameters $\Omega_M$ and
$\Omega_\Lambda$ are {\it independent} of the value for the Hubble
constant or the calibration of the SN Ia absolute magnitude.  

Indications for $\Omega_M$ and $\Omega_\Lambda$, independent from $H_0$,
can be found by reducing our three-dimensional PDF to
two dimensions.
A joint confidence region for $\Omega_M$ and $\Omega_\Lambda$ is
derived from our three dimensional likelihood space
\bq p(\Omega_M,\Omega_\Lambda \vert {\bf \mu_0})=\int_{-\infty}^\infty
p(\Omega_M,\Omega_\Lambda,H_0\vert {\bf \mu_0}) \, dH_0. \eq
The likelihood that the cosmological constant is greater than zero is
given by summing the likelihood for this region of parameter space, \bq P(\Omega_\Lambda > 0\vert {\bf \mu_0})=\int_0^\infty d\Omega_\Lambda
\int_0^\infty   p(\Omega_M,\Omega_\Lambda\vert {\bf \mu_0}) \, d\Omega_M. \eq This
integral was evaluated numerically over a wide and finely spaced grid
of cosmological
parameters for which equation (11) is non-trivial.  

 From the
nine spectroscopic high-redshift SNe Ia with well-observed light and
color curves,  a non-negligible positive cosmological constant is
strongly preferred at the 99.6\% (2.9$\sigma$) and $>$99.9\%
(3.9$\sigma$) confidence levels for the
MLCS and template fitting methods, respectively (see Table 8).  This region of
parameter space is nearly identical to the one which results in an
eternally expanding Universe.  Boundless expansion occurs for a cosmological
constant of 
\bq \Omega_\Lambda \geq \left[  \begin{array}{ll} 0 &  0 \leq \Omega_M \leq 1 \\
4\Omega_M\{\cos[{1 \over 3} \cos^{-1} ({1-\Omega_M \over \Omega_M}) +{4\pi \over 3}]\}^3 &  \Omega_M > 1  \end{array} \right] \eq
(Carroll, Press, \& Turner 1992), and its likelihood is
\bq \int_0^1 d\Omega_M \int_0^\infty p(\Omega_M,\Omega_\Lambda \vert {\bf \mu_0}) \, d\Omega_\Lambda + \int_1^\infty d\Omega_M
\int_{4\Omega_M\{\cos[{1 \over 3} \cos^{-1} ({1-\Omega_M \over \Omega_M}) +{4\pi \over 3}]\}^3}^\infty p(\Omega_M,\Omega_\Lambda\vert {\bf \mu_0}) \, d\Omega_\Lambda.\eq
The preference for eternal expansion is numerically equivalent to the
confidence levels cited for a non-negligible, positive cosmological constant.

We can include external constraints on $\Omega_M$, $\Omega_\Lambda$,
or their sum to further refine our determination of the cosmological parameters.
For a spatially flat Universe (i.e., $\Omega_M+\Omega_\Lambda \equiv \Omega_{tot} \equiv 1$), we find
  $\Omega_\Lambda=0.68 \pm 0.10$ ($\Omega_M=0.32 \pm 0.10$) and
$\Omega_\Lambda=0.84 \pm 0.09$ ($\Omega_M=0.16 \pm 0.09$) for MLCS and
  template fitting, respectively (see Table 8).  The hypothesis that matter provides the closure density (i.e., $\Omega_M=1$) is ruled out at the 7$\sigma$ to 9$\sigma$ level by either method. Again,
  $\Omega_\Lambda > 0$ and an eternally expanding Universe are strongly
  preferred, at this same confidence level.  We emphasize
  that these constraints reflect statistical errors only; systematic
  uncertainties are confronted in \S 5.

Other measurements based on the mass, light, x-ray emission, numbers,
and motions of clusters of galaxies provide constraints on the
mass density which have yielded typical values of 
$\Omega_M \approx 0.2-0.3$ (Carlberg et al. 1996; Bahcall, Fan, \& Cen 1997; Lin
et al. 1996; Strauss \& Willick 1995).  Using the constraint that
$\Omega_M \equiv 0.2$ provides
a significant indication for a cosmological constant: $\Omega_\Lambda=0.65
\pm 0.22$ and $\Omega_\Lambda=0.88 \pm 0.19$ for the MLCS and template
fitting methods, respectively (see Table 8).  For $\Omega_M \equiv
0.3$ we find $\Omega_\Lambda=0.80
\pm 0.22$ and $\Omega_\Lambda=0.96 \pm 0.20$ for the MLCS and template
fitting methods, respectively.

 If we instead demand that $\Omega_\Lambda \equiv 0$, we are forced to relax the
requirement that $\Omega_M \geq 0$ to locate a global minimum in our
$\chi^2$ statistic.  Doing so yields an
unphysical value of $\Omega_M=-0.38 \pm 0.22$ and $\Omega_M=-0.52 \pm
0.20$ for the MLCS and template fitting approaches, respectively (see Table 8).  This result emphasizes the
need for a positive cosmological constant for a plausible fit.

For the four sparsely observed SNe Ia (SN 1996R, SN 1996T, SN 1995ao,
and SN 1995ap), we employed the snapshot distance method (Riess et al.
1998a) to determine the luminosity
distances.    Unfortunately, the
low priority given to these objects resulted in observations not only
limited in frequency but in signal-to-noise ratio as well.  Consequently, these 4
distances are individually uncertain at the 0.4-0.6 mag level.  We
have compared these distances directly to a set of nine SNe Ia distances
measured by the same snapshot method with
$0.01 \leq z \leq 0.83$ from Riess et al. (1998a) and reprinted here in Tables 7 and 9.  This approach avoids
the requirement that distances calculated from light curves and the
snapshot method be on the same distance scale although this has been
shown to be true (Riess et al. 1998a).

  The complete but sparse set of
13 snapshot distances now including six SNe Ia
with $z \geq 0.16$ yields conclusions which are
less precise but fully consistent with the {\it statistically
independent} results from the well-sampled SN Ia
light curves (see Table 8).  

   Having derived the two PDFs, $p$($\Omega_M,\Omega_\Lambda$),
for the $\sim 40$ SNe Ia light curves and the 13 incomplete (``snapshot'')
SNe Ia light curves independently, we can multiply the two PDFs to yield the PDF
for all $\sim 50$ SNe Ia which includes 15 SNe with $0.16 \leq z \leq
0.62$.  Contours of constant PDF from the MLCS method and the template
fitting method, each combined with the snapshot PDF,
are shown in Figures 6 and 7.  These contours are closed by their
intersection with the line $\Omega_M=0$ and labeled by the total
probability contained within.  

   Including the snapshot distances modestly strengthens all of the
previous conclusions about the detection of a non-negligible, positive
cosmological constant (see Table 8).  This set of 15 high-redshift SNe Ia
favors $\Omega_\Lambda \geq 0$ and an eternally expanding Universe at
99.7\% (3.0$\sigma$) and
$>$99.9\% (4.0$\sigma$) confidence for the MLCS and template fitting
methods, respectively.  This complete set of {\it spectroscopic} SNe Ia represents the full
strength of the high-redshift sample and provides the most reliable results.

   A remarkably high-redshift supernova ($z=0.97$), SN 1997ck,  was
{\it excluded} from all these analyses due to its uncertain
extinction and the absence of a spectroscopic identification.
Nevertheless, if we assume a negligible extinction of $A_B=0.0 \pm
0.1$ for SN
1997ck as observed for the rest of our high-redshift sample and
further assume it is of type Ia, as its well-observed $B$ rest-frame
light curve suggests (see Figure 3), we could
include this object in our previous analysis (see Table 8).  As seen in Figures 6 and
7, SN 1997ck constrains specific values of $\Omega_M$ and $\Omega_\Lambda$
by effectively closing our confidence contours because of the increased
redshift range of this augmented sample.  The values implied using SN
1997ck and the rest of the spectroscopic SNe Ia, under the previous assumptions, are $\Omega_M=0.24^{+0.56}_{-0.24}$, $\Omega_\Lambda=0.72^{+0.72}_{-0.48}$ from the MLCS
method and $\Omega_M=0.80^{+0.40}_{-0.48}$, $\Omega_\Lambda=1.56^{+0.52}_{-0.70}$ from the
template fitting method.  The preference for a
non-negligible, positive cosmological constant remains strong (see
Table 8).  

As seen in Table 8, the values of the $\chi^2_{\nu}$ for
the cosmological fits are reassuringly close to unity.  This statement
is more meaningful for the MLCS distances which are accompanied by
statistically reliable estimates of the distance uncertainty (Riess, Press, \& Kirshner 1996a).
The values for $\chi^2_{\nu}$ indicate a good
agreement between the expected distance uncertainties and the observed
distance dispersions around the best fit model.  They
 leave little room for sources of additional variance, as might be
introduced
by a significant difference between the properties of SN Ia at high
and low redshift.

\subsection{Deceleration Parameter}

   An alternate approach to exploring the expansion history of the
Universe is to measure the current ($z=0$) deceleration parameter,
$q_0\equiv -\ddot a(t_0) a(t_0)/\dot a^2(t_0)$, where $a$ is the
cosmic scale factor. 
Because the deceleration is defined at
the current epoch and the supernovae in our sample cover a wide range
in redshift, we can only determine the value of $q_0$ within the
context of a model for its origin.
Nevertheless, for moderate values of deceleration (or acceleration) the determination of $q_0$
from our SNe, all of which are at $z < 1$,  provides a valuable
description of the current deceleration parameter valid for most
equations of state of the Universe.

   We have derived estimates of $q_0$ within a two-component model where
$q_0={\Omega_M \over 2} - \Omega_\Lambda$.  This definition assumes
that the only sources of the current deceleration are mass density and the cosmological constant.  A more complete
definition for $q_0$ would include all possible forms of energy density
(see Caldwell, Dave, \& Steinhardt 1998) but is beyond the scope of this
paper.  From our working definition of $q_0$, negative values for the
current deceleration (i.e., accelerations) are generated only by a
positive cosmological constant and not from unphysical, negative mass density.

   Current acceleration of the expansion occurs for a cosmological constant of
\bq \Omega_\Lambda \geq {\Omega_M \over 2}, \eq and its likelihood is
\bq P(q_0 < 0\vert {\bf \mu_0})=\int_0^\infty d\Omega_M \int_{\Omega_M \over 2}^\infty p(\Omega_M,\Omega_\Lambda\vert {\bf \mu_0}) \,
d\Omega_\Lambda \eq considering only $\Omega_M \geq 0$. Figures 6 and 7 show the boundary between
current acceleration and deceleration as well as lines of constant
$q_0$.  For the complete set of supernova distances (excluding SN
1997ck), current acceleration is strongly preferred at the 99.5\% (2.8$\sigma$)
confidence level for the MLCS method and $>$99.9\% level
(3.9$\sigma$) for the template fitting approach.  The
most likely value for $q_0$ is given by the peak of the distribution 
\bq p(q_0\vert {\bf \mu_0})= \int_0^\infty d\Omega_M
\int_{-\infty}^\infty 
p(\Omega_M,\Omega_\Lambda \vert q_0,{\bf \mu_0}) \, d\Omega_\Lambda. \eq  This
expression determines the likelihood of a given $q_0$ from the sum of
the likelihoods of the combinations of $\Omega_M$ and $\Omega_\Lambda$
which produce that value of $q_0$.  Values
for $q_0$ and their uncertainties for the different methods and sample cuts are summarized in
Table 8.  With the current sample we find a robust indication for the
sign of $q_0$ and a more uncertain estimate for its value, $q_0=-1.0 \pm 0.4$.  Because lines of
constant $q_0$ are skewed with respect to the major axis of our
uncertainty contours, more SNe Ia at redshifts greater than $z=0.5$
will be needed to yield a more robust indication for the value of $q_0$.
 
\subsection{Dynamical Age of the Universe} 

   The dynamical age of the Universe can be calculated from the cosmological parameters. In an empty Universe with no
cosmological constant, the dynamical age is simply the inverse of the
Hubble constant; there is no deceleration.  SNe Ia have been used to map the nearby Hubble flow resulting in a precise
determination of the Hubble constant (Hamuy et al. 1995, 1996a; Riess, Press, \& Kirshner 1995,
1996a).  For a more complex cosmology, integrating the
velocity of the expansion from the current epoch ($z=0$) to the beginning
($z=\infty$) yields an expression for the dynamical age
\bq t_0(H_0,\Omega_M,\Omega_\Lambda)= H_0^{-1} \int_0^\infty
(1+z)^{-1}[(1+z)^2(1+\Omega_Mz)-z(2+z)\Omega_\Lambda
]^{-1/2} \, dz  \eq (Carroll, Press, \& Turner 1992).  Combining a PDF for the
cosmological parameters, $p$(H$_0$,$\Omega_M$,$\Omega_\Lambda$), with the
above expression we can derive the PDF for the age of
the Universe:
\bq p(t_0\vert {\bf \mu_0})=\int_{-\infty}^\infty dH_0 \int_0^\infty d\Omega_M
\int_{-\infty}^\infty 
p(H_0,\Omega_M,\Omega_\Lambda \vert t_0,{\bf \mu_0}) \, d\Omega_\Lambda. \eq  
Equation (19) expresses the likelihood for a given age, $t_0$, as
the sum of the likelihoods of all combinations of $H_0$, $\Omega_M$,
and $\Omega_\Lambda$ which result in the given age.
The peak of this
function provides our maximum likelihood estimate for the dynamical
age, $t_0$.  Without SN 1997ck, the peak is at 13.6$^{+1.0}_{-0.8}$ Gyr from the MLCS PDF.
For the template fitting approach the peak occurs at 14.8$^{+1.0}_{-0.8}$  Gyr.
 A naive combination of the two distributions yields an
estimate of 14.2$^{+1.0}_{-0.8}$ Gyr adopting either method's uncertainty (see Figure 8).  Again, these
errors include only the statistical uncertainties of the measurement.
Including the {\it systematic} uncertainty of the Cepheid distance
scale, which may be as much as 10\%, a reasonable estimate of the dynamical age would be 14.2$\pm 1.5$ Gyr.

  An illuminating way to characterize the dynamical age independent of the Hubble constant is to measure the product $H_0 t_0$.  For the MLCS method, the template fitting method, and the combination of the two, we find $H_0 t_0$ to be 0.90, 0.96, and 0.93, respectively.  These values imply a substantially older Universe for a given value of $H_0$ in better accordance with globular cluster ages than the canonical value of $H_0 t_0$=2/3 for $\Omega_M=1$ and $\Omega_\Lambda=0$. Our determination of the dynamical age of the Universe is consistent
with the rather wide range of values of the ages using stellar theory or
radioactive dating.  Oswalt et al. (1996) have shown that the Galactic
disk has a lower age limit of 9.5 Gyr measured from the cooling
sequence of the white dwarfs.  The radioactive dating of stars via the
thorium and europium abundances gives a value of $15.2 \pm 3.7$ Gyr
(Cowan et al. 1997).  We can expect these ages to become more precise
as more objects are observed.

Perhaps the most widely quoted ages of the Universe come from the age
estimates of globular cluster stars. These are dependent on the
distance scale used and the stellar models employed. Vandenberg,
Stetson, \& Bolte (1996) note that these two effects generally work
in the opposite direction: for instance, if one increases the distance
to the Large Magellanic Cloud (LMC), the dynamical age of the Universe increases
 while the age based on the cluster ages decreases
(the main-sequence turnoff is brighter implying a younger
population). This means that there is only a limited range in
cosmological and stellar models that can bring the two ages into
concordance.

Prior to {\it Hipparcos}, typical age estimates based on the subdwarf
distance scale were greater than 15 Gyr for
cluster ages. Bolte
\& Hogan (1995) find $15.8 \pm 2.1$ Gyr for the ages of
the oldest clusters, while Chaboyer, Demarque, \& Sarajedini (1996) find a typical age of 18
Gyr for the oldest clusters. Chaboyer (1995) also estimates the full range of viable
 ages to be
11-21 Gyr with the dominant error due to uncertainties in the theory
of convection.  An independent distance scale based on parallaxes of
white dwarfs provides an age estimate for the globular cluster M4 of
$14.5-15.5$ Gyr (Renzini et al. 1996).

However, the {\it Hipparcos} parallaxes suggest an increased distance to
the LMC and the globular clusters (Feast \& Catchpole 1997; Reid
1997; but see Madore \& Freedman 1998). With this new distance scale, the ages of the clusters have
decreased to about 11.5 Gyr with an uncertainty of 2 Gyr (Gratton et al. 1997;
Chaboyer et al. 1998). Given the large range in ages from the
theoretical models of cluster turnoffs and the inconsistency of the
subdwarf and white dwarf distance scales applied to the ages of
globular clusters, a robust estimate for
the ages of the globular clusters remains elusive.
 Even with these uncertainties, the
dynamical age of the Universe derived here is consistent with 
the ages based on stellar theory or radioactive dating. Evidently,
there is no longer a problem that the age of the oldest stars is greater than
the dynamical age of the Universe.

   Despite our inability to place strong constraints on the values for
$\Omega_M$ and $\Omega_\Lambda$ independently, our experiment is
sensitive to the {\it difference} of these parameters.  Because
the dynamical age also varies approximately as the difference in
$\Omega_M$ and $\Omega_\Lambda$, our leverage on the determination of
the dynamical age is substantial.  This point can be illustrated with a
display of lines of constant dynamical age as a function of $\Omega_M$
and $\Omega_\Lambda$; comparing Figure 9 to Figures 6
and 7, we see that the semi-major axes of our error ellipses are roughly
parallel to the lines of constant dynamical age.  Figure 9 also
indicates why the most likely value for the dynamical age differs from the dynamical age derived for the most likely
values of $H_0$, $\Omega_M$, and $\Omega_\Lambda$.  For a fixed value
of the Hubble constant, younger dynamical ages span a larger region of
the ($\Omega_M$,$\Omega_\Lambda$) parameter space than older ages.  This shifts the most likely value for $t_0$ towards a younger age
and results in a ``tail'' in the distribution,  $p(t_0\vert {\bf \mu_0})$, extending
towards older ages.

\section{Discussion}

    The results of \S 4 suggest an
eternally expanding Universe which is accelerated by energy in the vacuum.
  Although these data do not provide independent constraints on $\Omega_M$ and
$\Omega_\Lambda$ to high precision without ancillary assumptions or
inclusion of a supernova with uncertain classification,
specific cosmological scenarios can still be tested without these requirements.

   {\it High-redshift SNe Ia are observed to be dimmer than expected in
 an empty Universe (i.e., $\Omega_M=0$) with no
cosmological constant.}  A cosmological explanation for this observation
is that a positive vacuum energy density accelerates the
expansion.  Mass density in the Universe exacerbates this problem,
 requiring even more vacuum energy.  For a Universe with $\Omega_M=0.2$, the MLCS and
 template fitting distances to the well-observed SNe are 0.25 and 0.28 mag farther on average than the
 prediction from $\Omega_\Lambda=0$.  The average MLCS and template
 fitting distances are still 0.18 and 0.23 mag farther than required
 for a 68.3\% (1$\sigma$) consistency for a Universe with
 $\Omega_M=0.2$ and without a cosmological constant.

Depending on the method used to
 measure all the spectroscopically confirmed SN Ia
distances, we find $\Omega_\Lambda$ to be inconsistent with zero at the
99.7\% (3.0$\sigma$) to $>$99.9\% (4.0$\sigma$) confidence level.
 Current acceleration of the expansion is preferred at the
99.5\% (2.8$\sigma$) to $>$99.9\% (3.9$\sigma$) confidence level.
The ultimate fate of the
Universe is sealed by a positive cosmological constant.  Without a
restoring force provided by a surprisingly large mass density
 (i.e., $\Omega_M > 1$) the Universe will continue to expand forever.

How reliable is this conclusion?  Although the statistical inference is strong, here we
explore systematic uncertainties in our results with special attention to those that can lead
to overestimates of the SNe Ia distances.

\subsection{Evolution}

   The local sample of SNe Ia displays a weak correlation between light
curve shape (or luminosity) and host galaxy type.  The sense of the
correlation is that the most luminous SNe Ia with the broadest light curves
only occur in late-type galaxies.  Both early-type and late-type galaxies
provide hosts for dimmer SNe Ia with narrower light curves (Hamuy et al.
1996c).  The mean luminosity difference for SNe Ia in late-type and
early-type galaxies is $\sim 0.3$ mag (Hamuy et al. 1996c).  In addition, the SN Ia rate per unit luminosity is almost
twice as high in late-type galaxies as in early-type galaxies at the
present epoch (Cappellaro et al. 1997).  This suggests that a population of
progenitors may exist in late-type galaxies which is younger and gives
rise to brighter SNe Ia (with broader light curves) than those contained in early-type galaxies or
within pockets of an older stellar population in the late-type galaxies.  Such observations could indicate an evolution
of SNe Ia with progenitor age.

H\"{o}flich, Thielemann,  \& Wheeler (1998) calculate differences in the
light curve shape, luminosity, and spectral characteristics of SNe Ia
as a function of 
the initial composition and metallicity of the white dwarf
progenitor.  As we observe more distant samples, we expect
the progenitors of SN Ia to come from a younger and more metal-poor population of stars.  H\"{o}flich, Thielemann, \& Wheeler (1998) have shown that a
reduction in progenitor metallicity by a factor of 3 has little effect on the
SN Ia bolometric luminosity at maximum.  For their models, such a change in
metallicity can alter the peak luminosity by
small amounts ($\sim 0.05$ mag) in rest-frame $B$ and $V$, accompanied
by detectable spectral signatures.  These spectral indicators of
evolution are expected to be most discernible in the rest-frame $U$
passband where line blanketing is prevalent.  Future detailed spectral
 analyses at these short wavelengths might provide a constraint on a
variation in progenitor metallicity.

The effect of a decrease in SN Ia progenitor age at high redshift is predicted to be
 more significant than metallicity (H\"{o}flich, Thielemann, \& Wheeler 1998).
 Younger white dwarfs are expected to evolve from
 more massive stars with a lower ratio of C/O in their cores.
The
lower C/O ratio of the white dwarf reduces the amount of $^{56}$Ni synthesized in the
 explosion, but an anticipated slower rise to maximum conserves
 more energy for an increased maximum brightness.
 By reducing the C/O ratio from 1/1 to 2/3, the $B-V$ color at maximum is expected to become redder by
 0.02 mag and the post-maximum decline would become steeper.  This prediction of a brighter SN Ia
 exhibiting a faster post-maximum decline is opposite to what is seen
 in the nearby sample (Phillips 1993; Hamuy et al. 1995; Hamuy et
 al. 1996a,b,c,d; Riess, Press, \& Kirshner 1996a; Appendix) and will be readily testable
for an enlarged high redshift sample.  Specifically, 
a larger sample 
of distant SNe Ia (currently being compiled) would allow us to
 determine the light curve shape relations at high-redshift and test
whether these evolve with look-back time.  Presently, our sample
is to small to make such a test meaningful.

We expect that the relation between light curve shape and luminosity that
applies to the range of stellar populations and progenitor ages encountered in the late-type and
early-type hosts in our nearby sample should also
be applicable to the range we encounter in our distant sample.
In fact, the range of age for SN Ia progenitors in
the nearby sample is likely to be {\it larger} than the change in mean progenitor age over
 the 4 to 6 Gyr look-back time to the high-redshift sample.  
Thus, to first order at least, our local sample should correct our
distances for progenitor or age effects.

We can place empirical constraints on the effect that a change in the
progenitor age would have on our SN Ia distances by
comparing subsamples of low redshift SNe Ia believed to arise from old and 
young progenitors.  In the nearby
sample, the mean difference between the distances for the early-type
(8 SNe Ia)
and late-type hosts (19 SNe Ia), at a given redshift, 
is 0.04 $\pm$ 0.07 mag from the MLCS method.  This difference
is consistent with zero.  Even if
the SN Ia progenitors evolved from one 
population at low redshift to the other at high
redshift, we still would not explain the surplus in mean distance of 0.25 mag over the $\Omega_\Lambda=0$ prediction.
For the template fitting approach, the mean difference in distance for
SNe Ia in early-type and
late-type hosts is 0.05 $\pm$ 0.07 mag.  Again, evolution
provides an inadequate explanation for the 0.28 mag
difference in the template fitting SNe Ia distances and the $\Omega_\Lambda=0$ prediction.  

However, the low-redshift sample is dominated by late-type hosts
and these may contain a number of
older progenitors.  It is therefore 
difficult to assess the precise effect of a decrease in progenitor age at 
high redshift from the consistency of distances to early-type and late-type
hosts (see Schmidt et al. 1998).  If, however, we believed that young progenitors give rise to
brighter SNe Ia with broader light curves (Hamuy et al. 1996c) as discussed above, we could more directly determine the effect on distance determinations
of drawing our high-redshift sample from an increasingly youthful population of progenitors.  The mean difference in
the Hubble line defined by the full nearby sample and the subsample of SNe Ia
with broader than typical light curves ($\Delta < 0$) is 0.02 $\pm$ 0.07 for the MLCS method.  For the template fitting method, the difference between
the full sample and those with broader light curves ($\Delta m_{15}(B) < 1.1$) is 0.07 $\pm$ 0.07.  Again, we find no indication of a systematic change in 
our distance estimates with a property that may correspond to a decrease in progenitor age.  Another valuable 
test would be to compare low-redshift distances to starburst and irregular
type galaxies which presumably are hosts to progenitors which are
young and metal-poor.  Such a nearby sample may yield the closest
approximation to the SNe Ia observed at high redshift.
Future work will be needed to gather this informative sample which
would be composed of objects such as SN 1972E in NGC 5253 (which we 
does fit the luminosity light curve shape relations; Hamuy et al 1996b).     

Another check on evolutionary effects is to test whether the
distribution of light curve decline rates is similar between
the nearby sample of supernovae and the high-redshift sample. 
 Figure 10
shows the observed distribution of the MLCS light-curve shape
parameters, $\Delta$, and the template fitting parameters, $\Delta
m_{15}(B)$, with redshift.
A Kolmogorov-Smirnov test shows no significant difference in the
distributions of the low and high-redshift
samples, but the sample is too small to be statistically
significant. The actual difference in mean luminosity between the low-redshift and high-redshift samples implied by the light curve
shapes is 0.02 mag by either method.   We conclude that there is no obvious difference between the shapes
of SNe~Ia light curves at $z \approx 0$ and at $z \approx 0.5$.

 It is reassuring that initial comparisons of high-redshift SN Ia
spectra appear remarkably similar to those observed at low-redshift.
This can be seen in the high signal-to-noise ratio spectra of SN
1995ao ($z=0.30$) and SN 1995ap ($z=0.23$) in Figure 1. Another
demonstration of this similarity at even higher redshift
is shown in Figure 11 for SN 1998ai ($z=0.49$; IAUC 6861) whose
light curve was not used in this work.  The spectrum of SN 1998ai was
obtained at the Keck telescope with a 5 x 1800 s exposure using LRIS and
was reduced as described in \S 2.2 (Filippenko et al. 1998).
The spectral characteristics of this SN Ia appear to be indistinguishable
from the range of characteristics at low redshift to good precision.
In additon, a time sequence
of spectra of SN Ia
1997ex ($z$=0.36; Nugent et al. 1998a) compared with those of local SNe
Ia reveals
no significant spectral differences (Filippenko et al. 1998).

We expect that our local
calibration will work well at eliminating any pernicious drift in the
supernova distances between the local and distant samples.
Until we know more about the
stellar ancestors of SNe Ia, we need to be vigilant for changes in
the properties of the supernovae at significant look-back times.  
Our distance measurements could be particularly sensitive to changes in
the colors of SNe Ia for a given light curve shape.  Although
our current observations reveal no indication of evolution
of SNe Ia at $z \approx 0.5$, evolution remains a serious concern which can
only be eased and perhaps understood by future studies.

\subsection{Extinction}
 Our SNe Ia distances have the important advantage of including corrections for
interstellar extinction occurring in the host galaxy and the Milky
Way.  The uncertainty in the extinctions is a significant component of
error in our distance uncertainties.  Extinction corrections based on
the relation between SN Ia colors and luminosity improve distance precision for a
sample of SNe Ia that includes objects with substantial extinction
(Riess, Press, \& Kirshner 1996a).
Yet, in practice, we have found negligible extinction to the high-redshift
SNe Ia.  The mean $B-V$ color at maximum is $-0.13 \pm 0.05$ from the MLCS method and $-0.07 \pm 0.05$ from the template fitting approach, consistent with an unreddened $B-V$ color of $-0.10$ to $-0.05$ expected for slowly declining light curves as observed in the high-redshift sample (Riess, Press, \& Kirshner 1996a; Appendix).

Further, the consistency of
the measured Hubble flow from SNe Ia with late-type and early-type hosts (\S 5.1)
shows that the extinction corrections applied to dusty SNe Ia at low redshift do not alter the
expansion rate from its value measured from
SNe Ia in low dust environments.  The conclusions reached in \S 4
would not alter if low and high-redshift SNe with significant extinction
were discarded rather than included after a correction for extinction.

The results of \S 4 do not depend on the value of the ratios between
color excess and selective absorption used to determine the
extinctions of the high-redshift sample because the mean observed
reddening is negligible.   Some modest departures from the Galactic
reddening ratios have been observed in the Small and Large Magellanic
Clouds, M31, and the Galaxy, and they have been linked to metallicity
variations (Walterbos 1986; Hodge \& Kennicutt 1982; Bouchet et al
1985; Savage \& Mathis 1979).  Although our current understanding of
the reddening ratios of interstellar dust at high redshift is limited,
the lack of any significant color excess observed in the high-redshift sample indicates that the type of interstellar dust which reddens optical light is not obscuring our view of these objects.

Riess, Press, \& Kirshner
(1996b) found indications that the Galactic ratios between selective absorption
and color excess are similar for host galaxies in the nearby ($z$ $\leq
0.1$) Hubble flow.  Yet, what if these ratios changed with look-back time?  Could an
evolution in dust grain size descending from ancestral interstellar ``pebbles'' at
higher redshifts cause us to underestimate the extinction?  Large dust
grains would not imprint the reddening signature of typical
interstellar extinction upon which our corrections rely.
However, viewing our SNe through such grey interstellar grains would also induce
a {\it dispersion} in the derived distances.  To estimate the size of the dispersion,
we assume that the grey extinction is distributed in
galaxies in the same way as typical interstellar extinction.

  Hatano,
Branch, \& Deaton (1997) have calculated the expected distribution of
SN Ia extinction
along random lines of sight in the host galaxies.
A grey extinction distribution similar to theirs could yield differing amounts of mean grey extinction
depending on the likelihood assigned to observing an extinction of $A_B$=0.0
mag.  In the following calculations we vary only the likelihood of
$A_B$=0.0 mag to derive new extinction distributions with varying means.
These different distributions also have differing dispersions of extinction.
A mean grey extinction of 0.25 mag would be required
 to explain the
measured MLCS distances without a cosmological constant.  Yet the dispersion of individual extinctions
for a distribution with a mean of 0.25 mag 
would be $\sigma_{A_B}$=0.40 mag, significantly {\it larger} than the 0.21
mag dispersion observed in the high-redshift MLCS distances.
Grey extinction is an even less
likely culprit with the template fitting approach; a distribution with a mean grey extinction of 0.28
mag, needed to replace a cosmological constant, would yield a
dispersion of 0.42 mag, significantly higher than the
distance dispersion of 0.17 mag observed in the high-redshift template fitting
distances.

 Furthermore,
most of the observed scatter is already consistent with the estimated
statistical errors as evidenced by the $\chi^2_\nu$ (Table 8), leaving little to be caused by grey
extinction.
Nevertheless, if we assumed that {\it all} of the observed scatter were due to grey
extinction,
the mean shift in the SNe Ia distances would only be 0.05 mag.
With the observations presented here, we cannot rule out this modest
amount of grey interstellar extinction.

  This argument applies not only to exotic grey extinction but to
 any interstellar
 extinction not accounted for which obscures SNe Ia.  Any spotty interstellar
   extinction which varies with
   line-of-sight in a way similar to the Hatano,
Branch, \& Deaton (1997) model of galaxies
   will add dispersion to the SN Ia distances.  The low
 dispersion measured for the high-redshift sample places a strong
 limit on any mean spotty interstellar extinction.

  Grey intergalactic extinction could dim the SNe without either
telltale reddening or dispersion, if all lines of sight to a given redshift
had a similar column density of absorbing material.
The component of the intergalactic medium with such uniform coverage
corresponds to the gas clouds producing Lyman-$\alpha$
forest absorption at low redshifts.  These clouds have
individual H I column densities less than about
$10^{15} \, {\rm cm^{-2}}$ (Bahcall et al. 1996).  However, these
 clouds display low metallicities, typically less than 10\% of
solar. Grey extinction would require larger dust grains
which would need a larger mass in heavy elements than typical
interstellar grain size distributions to achieve a given extinction.  Furthermore, these
clouds reside in hard radiation environments hostile
to the survival of dust grains.  Finally, the existence of grey intergalactic extinction would only augment the
already surprising excess of galaxies in high-redshift galaxy surveys
(Huang et al. 1997).

  We conclude that  grey extinction does not
seem to provide an observationally or physically plausible explanation
for the observed faintness of high-redshift SNe Ia.

\subsection{Selection Bias}

Sample selection has the potential to distort the comparison of
nearby and distant supernovae. Most of our nearby ($z < 0.1$) sample of SNe~Ia
was gathered from the Cal\'an/Tololo survey (Hamuy et al. 1993a) which employed the blinking of photographic plates obtained at
different epochs with Schmidt telescopes and
from less well-defined searches (Riess et al. 1998b).
Our distant ($z>0.16$) sample was obtained by
subtracting digital CCD images at different epochs with the same instrument setup. 

If they were limited by the flux of the detected events, both nearby and distant SN Ia searches would
preferentially select
intrinsically luminous objects because of the larger volume of space in which these objects can
be detected.  This well-understood selection effect could be further complicated by the properties
of SNe Ia; more luminous supernovae have
broader light curves (Phillips 1993; Hamuy et al. 1995, 1996c; Riess, Press, \& Kirshner
1995, 1996a).  The brighter supernovae remain above a detection limit
longer than their fainter siblings, yet also can fail to rise above the
detection limit in the time interval between successive search
epochs.   The complex process by which SNe Ia are selected in low and
high-redshift searches can be best understood with simulations (Hamuy
\& Pinto 1998).  Although selection effects could alter the ratio of
intrinsically dim to bright SNe Ia in our samples relative to the true population, our use of the light curve shape to
determine the supernova's luminosity should correct most of
this selection bias on our distance estimates.  
However, even after our light-curve shape correction, SNe~Ia still have a small dispersion as distance
indicators ($\sigma \approx 0.15$ mag), and any search program would
still preferentially select objects which are brighter than average
for a particular light curve shape and possibly select objects whose light curve shapes
aid detection.

To investigate the consequence of sample selection effects, we
used a Monte Carlo simulation to understand how SNe Ia in our nearby and
distant samples were chosen. For the purpose of this simulation we
first assumed that the SN Ia rate is constant with look-back time.  We
assembled a population of SNe Ia with luminosities described by a
Gaussian random variable $\sigma_{M_B}=0.4$ mag and light-curve shapes
which correspond to these luminosities as described by the MLCS vectors
(see the Appendix).  A Gaussian random uncertainty of $\sigma = 0.15$ mag is assumed in the determination of
absolute magnitude from the shape of a supernova's light curve.  The time interval
between successive search epochs, the search epoch's limiting
magnitudes, and the apparent light-curve shapes were used to determine which
SNe Ia were ``discovered'' and included in the simulation sample.  
A separate simulation was used to select nearby objects, with the
appropriate time interval between epochs and estimates of limiting
magnitudes.  The results are extremely encouraging, with recovered
values exceeding the simulated value of $\Omega_M$ or $\Omega_\Lambda$
by only 0.02 for these two parameters considered separately.
Smoothly increasing the SN Ia rate by a factor of 10 by $z=1$ doubles
this bias to 0.04 for either parameter.
There are two reasons we find such a
small selection bias in the recovered cosmological parameters.  First,
the small dispersion of our distance indicator results in only a modest
selection bias.  Second, both
nearby and distant samples include an excess of brighter than average
SNe, so the {\it difference} in their individual selection biases
remains small.

As discussed by Schmidt et al. (1998), obtaining accurate limiting magnitudes is complex
for the CCD-based searches, and essentially impossible for the photographic searches.
Limiting magnitudes vary from frame to frame, night to night,
and film to film, so it is
difficult to use the actual detection limits in our simulation.
Nevertheless, we have run simulations varying the limiting magnitude, and this
does not change the results significantly.  We have also tried
increasing the dispersion in 
the SN~Ia  light curve shape vs. absolute magnitude correlation
at wavelengths shorter than $5000$~\AA. Even doubling the distance dispersion
of SNe~Ia (as may be the case for rest-frame $U$) does
not significantly change the simulation results.

Although these simulations bode well for using SNe~Ia to measure
cosmological parameters,
there are other differences between the way nearby and distant
supernova samples are selected which are more difficult to model and
are not included in our
present simulations.  Von Hippel, Bothun, \& Schommer (1997) have shown that the selection
function of the nearby searches is not consistent with that of a
strict magnitude-limited search. It is unclear whether a photographic
search selects SNe Ia with different parameters or environments
than a CCD search or how this could affect a comparison of samples.
   Future work on quantifying the
selection criteria of the samples is needed.
A CCD search for SNe Ia in Abell clusters by Reiss et al. (1998) will
soon provide a nearby SN Ia sample with better understood selection criteria.  
Although indications from the distributions of SN Ia parameters
suggest that both our searches have sampled the same underlying population
(see Figure 10), we must continue to
be wary of subtle selection effects which might bias the comparison of
SNe~Ia near and far.

\subsection{Effect of a Local Void}

   It has been noted by Zehavi et al. (1998) that the 
SNe Ia out to 7000 km s$^{-1}$ exhibit an expansion rate
which is 6\% greater than that measured for
 the more distant objects.  The
significance of this peculiar monopole is at the 2$\sigma$ to
3$\sigma$ confidence level; it is not inconsistent with the upper limit
of $\sim$ 10\% for the difference between the local and global values
of $H_0$ found by Kim et al. (1997).  The implication is that
the volume out to this distance is underdense relative to the global
mean density.  This effect appears as an excess redshift for a given
distance modulus (within 7000 km s$^{-1}$) and can be seen with both the MLCS method and
the template fitting method in Figures 4 and 5 .  

   If true, what effect would this result have on our conclusions?
In principle, a local void would increase the expansion rate
measured for our low-redshift sample relative to the true, global
expansion rate.  Mistaking this inflated rate for the
global value would give the false impression of an increase
in the low-redshift expansion rate relative to the high-redshift
expansion rate.  This outcome could be incorrectly attributed to the influence
of a positive cosmological constant. In practice, only a small
fraction of our nearby sample is within this local void,
reducing its effect on the determination of the low-redshift expansion rate.

As a test of the effect of a local void on our constraints for the cosmological
parameters, we reanalyzed the data discarding the seven SNe Ia within 7000 km s$^{-1}$ (108 Mpc for $H_0=65$).  The result was a reduction in the confidence that
$\Omega_\Lambda > 0$ from 99.7\% (3.0$\sigma$) to 98.3\% (2.4$\sigma$)
for the MLCS method and from
$>$99.9\% (4.0$\sigma$) to 99.8\% (3.1$\sigma$) for the template
fitting approach.  The tests for both methods excluded the
unclassified SN 1997ck and included the snapshot sample, the latter
without two SNe Ia within 7000 km s$^{-1}$.  As expected, the influence
of a possible local void on our cosmological
conclusions is relatively small.

\subsection{Weak Gravitational Lensing}

  The magnification and demagnification of light by large-scale
 structure can alter the observed magnitudes of high-redshift
supernovae (Kantowski, Vaughan, \& Branch 1995).  The effect of weak
 gravitational lensing 
on our analysis has been quantified by
Wambsganss et al. (1997) and summarized by Schmidt et al. (1998).  SN Ia light will, on
average, be demagnified by $0.5$\% at $z=0.5$ and $1$\% at $z=1$ in a Universe with a non-negligible cosmological constant.
Although the sign of the effect is the same as the influence of a
cosmological constant, the size of the effect is
negligible.  

Holz \& Wald (1997) have calculated the weak lensing effects on
supernova light from ordinary matter which is not smoothly distributed in
galaxies but rather clumped into stars (i.e., dark matter contained in MACHOS).
With this scenario, microlensing by compact masses becomes a more
important effect further decreasing the observed supernova
luminosities at $z=0.5$ by
0.02 mag for $\Omega_M$=0.2 (Holz 1998).  Even if most ordinary matter were
contained in compact objects, this effect would not be large enough to
reconcile the SNe Ia distances with the influence of ordinary matter alone.

\subsection{Light Curve Fitting Method}

   As described in \S 3.2, two different light curve fitting methods,
MLCS (Riess, Press, \& Kirshner 1996a; Appendix) and a template fitting approach (Hamuy et al. 1995, 1996d),
were employed to determine the distances to the nearby and
high-redshift samples.  Both methods use relations between light curve
shape and luminosity as determined from SNe Ia in the nearby
Hubble flow.  Both methods employ an extinction correction from the
measured color excess using relations between intrinsic color and
light curve shape.  In addition, both the MLCS and template fitting
methods yield highly consistent measurements for the Hubble constant of
 $H_0$=65.2 $\pm 1.3$ and $H_0$=63.8 $\pm 1.3$, respectively not including
 any uncertainty in the determination of the SN Ia absolute magnitude
which is the dominant uncertainty.  It is also worth noting that both
methods yield SN Ia distance dispersions of $\sim$ 0.15 mag when complete light curves in $B,V,R$, and $I$ are employed.  For the purpose of comparing
the same data at high and low redshifts, the use of SN Ia observations at low redshift were restricted to only $B$ and $V$ within 40 days of maximum light.

  Although the conclusions reached by the two methods when applied to the
high-redshift SNe are highly consistent, some differences are worth noting.
  There are small differences in the distance
predictions at high redshift.  For the distant sample, the template
fitting distances exhibit a scatter of 0.17 mag around the best fit
model as compared to 0.21 mag for the MLCS method.  In addition, the template fitting
distances to the high-redshift SNe Ia are on (weighted) average 0.03 mag farther than the MLCS
distances relative to the low-redshift sample.  These differences together result in slightly
different confidence intervals for the
two methods (see Figures 6, 7, and 8 and Table 8).  For the set of 10
well-observed SNe Ia, a sample with scatter 0.17 mag or less
is drawn from a population of scatter 0.21 mag 25\% of the time.
The chance that 10 objects could be drawn from this same population
with a mean difference of 0.03 mag is 66\%.   Future samples of SNe Ia will
reveal if the observed differences are explained by chance.  Until then,
we must consider the difference between the cosmological constraints reached from the two
fitting methods to be a systematic uncertainty.  Yet, for the data
considered here, both distance fitting methods unanimously favor
the existence of a non-negligible, positive cosmological constant and
an accelerating Universe.

\subsection{Sample Contamination}

   The mean brightness of SNe Ia is typically 4 to 40 times greater
   than that of any other type of supernova, favoring their detection
   in the volume of space searched at high redshift. Yet in the course
   of our high-redshift supernova search (and that of the Supernova
   Cosmology Project; Perlmutter et al. 1995) a small minority of
   other supernova types have been found and we must be careful
   not to include such objects in our SN Ia sample.   The classification of a supernova is determined from the presence or
   absence of specific features in the spectrum (Wheeler \& Harkness
   1990; Branch, Fisher, \& Nugent 1993; Filippenko 1997).  The
   spectra of Type Ia supernovae show broad Si II absorption near 6150
   \AA\, Ca II (H\&K) absorption near 3800 \AA\, a S II absorption
   doublet near 5300 \AA\ and 5500 \AA\, and numerous other
   absorption features with ionized Fe a major contributor (Filippenko 1997).
     For supernovae at high redshift, some of these characteristic
   features shift out of the observer's frequency range as other,
   shorter wavelength features become visible.  Classification
   is further complicated by low signal-to-noise ratio in
   the spectra of distant objects.
   The spectra of SNe Ia evolve with time along a remarkably
   reliable sequence (Riess et al. 1997). Final spectral
   classification is optimized by 
   comparing the observed spectrum to well-observed spectra of SNe Ia at the same
   age as determined from the light curves.

     For most of the spectra in Figure 1, the identification as a SN Ia
is unambiguous.  However, in three of the lowest signal-to-noise ratio
cases --
1996E, 1996H, and 1996I -- the wavelengths near Si II absorption (rest-frame 6150 \AA\ )  
were poorly observed and their classification warrants closer
scrutiny.  These spectra are inconsistent with Type II spectra
which show H$\beta$ (4861 \AA\ ) in emission and absorption and lack
Fe II features shortly after maximum.  These spectra are also
inconsistent with Type Ib spectra which would display He I~$\lambda$5876 absorption at
a rest wavelength of $\sim$ 5700 \AA .

  The most likely supernova type to be misconstrued as a Type Ia is a
Type Ic, as this type comes closest to matching the SN Ia spectral
characteristics. Although SN Ic spectra lack Si II and S II
absorption, the maximum-light spectra at 
blue wavelengths can resemble those of SNe Ia $\sim$ 2 weeks past
maximum when both are dominated by absorption lines of Fe II with P
Cygni profiles.  Type Ic events are rare and one luminous enough to be found in our search
would be rare indeed, but not without precedent.  An example
of such an object is SN 1992ar (Clocchiatti et al. 1998), which was
discovered in the course of the Cal\'{a}n/Tololo SN survey and which reached
an absolute magnitude, uncorrected for host galaxy dust extinction, of
$M_V = -19.3$ ($H_0$ = 65 km s$^{-1}$ Mpc$^{-1}$).  For both SN 1996H and SN 1996I,
 the spectral match with a Type Ia at rest wavelengths less
than 4500 \AA\ is superior to the fit to a Type Ic spectrum (see Figure 1).  In both
cases the spectra rise from deep troughs at the 3800 \AA\ Ca II break
(rest-frame)
to strong peaks at 3900 to 4100 \AA\ (rest-frame) as observed in SNe Ia.  Type
Ic spectra, by comparison, tend to exhibit a much weaker transition from trough to
peak redward of the Ca II break (see Figure 12).

For SN 1996E, the spectral
coverage does not extend blueward of a 
rest wavelength of 4225 \AA\ rendering this diagnostic unusable. The absence of pre-maximum observations of SN
1996E makes it difficult to determine the age of the spectrum and that
of the appropriate comparison spectra.
As shown in Figure 12,
the spectroscopic and photometric data for SN 1996E are consistent with a SN Ia
caught $\sim$ 1 week after maximum light, or a luminous SN Ic
discovered at maximum.  
There is a weak indication of S II absorption at $\sim$ 5375 \AA\
which favors classification as a Type Ia (see Figures 1 and 12), but
this alone does not provide a secure classification.
   Note that the $K$-corrections for a SN Ia or SN Ic at this redshift ($z=0.43$) would be nearly identical due to the excellent match of the observed filters ($B45$ and $V45$) to the rest-frame ($B$ and $V$) filters.

   We have reanalyzed the cosmological parameters discarding SN 1996E
   as a safeguard against the possible contamination of our
   high-redshift sample.  We also excluded SN 1997ck which, for lack of
   a definitive spectral classification, is an additional threat to
   contamination of our sample.  With the remaining
   ``high-confidence'' sample of 14 SNe Ia we find the statistical
   likelihood of a positive cosmological constant to be 99.8\% (3.1
   $\sigma$) from
   the MLCS method, a modest increase from 99.7\% (3.0
   $\sigma$) confidence when SN
   1996E is included.  For the template fitting approach, the
   statistical confidence in a positive cosmological constant remains
   high at $>$99.9\% (4.0 $\sigma$), the same result as with SN 1996E.
   We conclude that for this sample our results are robust against
   sample contamination, but the possible contamination of future
   samples remains a concern. Even given existing detector technology,
   more secure supernova classifications can be achieved with greater
   signal-to-noise ratios for observed spectra, with optimally timed search
   epochs which increase the likelihood of pre-maximum discovery, and with an
   improved empirical understanding of the differences among the
   spectra of supernova types.

\subsection{Comparisons}

The results reported here are consistent with other reported
observations of high-redshift SNe Ia from the High-z Supernova Search
Team (Garnavich et al. 1998;
Schmidt et al. 1998), and the improved statistics of this larger sample
reveal the potential influence of a positive cosmological constant.  

     These results are inconsistent at the $\sim 2\sigma$ confidence
level with those of Perlmutter et
al. (1997), who found $\Omega_M=0.94 \pm 0.3 \, (\Omega_\Lambda=0.06)$ for a flat Universe and $\Omega_M=0.88 \pm 0.64$
for $\Omega_\Lambda \equiv 0$.  They are marginally
consistent with those of Perlmutter et al. (1998) who, with the
addition of one very high redshift SN Ia ($z=0.83$), found
$\Omega_M=0.6 \pm 0.2 \, (\Omega_\Lambda=0.4)$ for a flat Universe and
$\Omega_M=0.2 \pm 0.4$ for $\Omega_\Lambda \equiv 0$. 

  Although the
experiment reported here is very similar to that performed by
Perlmutter et al. (1997, 1998), there are some differences worth
noting.  Schmidt et al. (1998), Garnavich et al. (1998), and this paper
explicitly correct for the effects of extinction evidenced by reddening of the
SNe Ia colors.  Not correcting for extinction in the
nearby and distant sample could affect the cosmological results in
either direction since we do not know the sign of the difference of
the mean extinction.
  In practice we have found few of
the high-redshift SNe Ia to suffer measurable reddening.  A number of objects in the
nearby sample display moderate extinction for which we make
individual corrections.  We also include the Hubble constant
as a free parameter in each of our fits to the other cosmological parameters.
Treating the nearby sample in the same way as the distant sample is a
crucial requirement of this work.  Our experience observing the
nearby sample aids our ability to accomplish this goal.

 The statistics of gravitational lenses provide an
alternate method for constraining the cosmological constant (Turner
1990; Fukugita, Futamase, \& Kasai 1990).  Although current
gravitational lensing limits for
the cosmological constant in a flat Universe ($\Omega_\Lambda \leq 0.66$ at 95\% confidence; Kochanek 1996) are not inconsistent with these results, they are
uncomfortably close.  Future analysis which seeks to limit systematic
uncertainties affecting both experiments should yield meaningful comparisons. 
The most incisive independent test may come from measurements of the
fluctuation spectrum of the cosmic microwave background.  While the supernova
measurements provide a good constraint on $\Omega_M - \Omega_\Lambda$,
the CMB measurements of the angular scale for the first Doppler peak, 
referring to much earlier epochs, are good measures of $\Omega_M + \Omega_\Lambda$ (White \& Scott 1996).  Since these constraints
are nearly orthogonal in the coordinates of Figure 6 and 7, the region of intersection
could be well defined.  Ongoing experiments from balloons and the South Pole
may provide the first clues to the location of where that intersection. 

 {\it Our detection of a cosmological constant is not limited by
statistical errors but by systematic ones.}  Further intensive study of
SNe Ia at low ($z$ $<$ 0.1), intermediate ($0.1 \leq z \leq 0.3$), and
high ($z >$ 0.3) redshifts is needed to uncover and quantify lingering
systematic uncertainties in this striking result.

\section {Conclusions}

1. We find the luminosity distances to well-observed SNe with 0.16 $\leq$ $z$ $\leq$ 0.97 measured by two methods to
be in excess of the prediction of a low mass-density ($\Omega_M$ $\approx
0.2$) Universe by 0.25
to 0.28 mag.  A cosmological explanation is provided by a
positive cosmological constant with 99.7\% (3.0$\sigma$) to $>$99.9\% (4.0$\sigma$)
confidence using the complete spectroscopic SN Ia sample and the prior
belief that $\Omega_M \geq 0$.

2. The distances to the spectroscopic sample of SNe Ia measured by two methods are consistent with a currently accelerating expansion
($q_0 \leq 0$) at the 99.5\% (2.8$\sigma$) to $>$99.9\% (3.9$\sigma$) level for
$q_0 \equiv {\Omega_M \over
2}-\Omega_\Lambda$ using the prior that $\Omega_M \geq 0$.  

3. The data favor eternal expansion as the fate of the Universe at
the 99.7\% (3.0$\sigma$) to $>$99.9\% (4.0$\sigma$) confidence level from the
spectroscopic SN Ia sample and the prior that $\Omega_M \geq 0$.

4. We estimate the dynamical age of the Universe to be 14.2 $\pm 1.5$
Gyr including systematic uncertainties, but subject to the zeropoint of
the current Cepheid distance scale used for the host galaxies of 
three nearby SNe Ia (Saha et al. 1994, 1997).    

5. These conclusions do not depend on inclusion of SN 1997ck ($z$=0.97),
whose spectroscopic classification remains uncertain, nor on which of two
light-curve fitting methods is used to determine the SN Ia distances.

6. The systematic uncertainties presented by grey
extinction, sample selection bias, evolution, a local void, weak
gravitational lensing, and sample contamination currently do not provide a
convincing substitute for a positive cosmological constant.  Further
studies are needed to determine the possible influence of any
remaining systematic uncertainties.

\bigskip
\bigskip 

   We wish to thank Alex Athey and S. Elizabeth Turner for their help
in the supernova search at CTIO.  We have benefited from helpful
discussions with Peter Nugent, Alex Kim, Gordon Squires, and Marc Davis and from the efforts of Alan Dressler, Aaron Barth, Doug
Leonard, Tom Matheson, Ed Moran, and Di Harmer.  The work at U.C. Berkeley was supported by the Miller Institute for Basic Research
in Science, by NSF grant AST-9417213 and by grant GO-7505 from
the Space Telescope Science Institute, which is operated by the
Association of Universities for Research in Astronomy, Inc., under
NASA contract NAS5-26555.  Support for AC was provided by
the National Science Foundation through grant \#GF-1001-95 from AURA, Inc.,
under NSF cooperative agreement AST-8947990 and AST-9617036, and from Fundaci\'on
Antorchas Argentina under project A-13313.  This work was supported at
Harvard University through NSF grants AST-9221648, AST-9528899, and an NSF
Graduate Research Fellowship.  CS acknowledges the generous support of the Packard Foundation and
the Seaver Institute.  This research was based in part on spectroscopic observations obtained with 
the Multiple Mirror Telescope, a facility operated jointly by the Smithsonian 
Institution and the University of Arizona.

\appendix 

\section {Appendix: MLCS}

     Following the success of Phillips (1993), Riess, Press, \& Kirshner (1995) employed
 a linear estimation algorithm (Rybicki \& Press 1992) to determine
the relationship between the shape of a SN Ia light curve and its 
peak luminosity.  This method was extended (Riess, Press, \& Kirshner 1996a) to utilize
the SN Ia color curves to quantify the amount of reddening by
interstellar extinction.  In this Appendix we describe further
refinements and optimization of the MLCS method for the application
to high-redshift SNe Ia.

   Previously, the MLCS relations were derived from a set similar to the nearby ($cz$ $\leq 2000 $ km s$^{-1}$) sample of Phillips (1993) (Riess, Press, \& Kirshner 1995, 1996a).  The relative luminosities of this ``training set'' of
SNe Ia were calibrated with independent distance indicators (Tonry
1991; Pierce 1994).  The absolute SN Ia luminosities were measured
from Cepheid variables populating the host galaxies (Saha et al.
1994, 1997).  Yet at moderate distances, the most reliable distance indicator available in nature
is the redshift.  The recent harvest of SN Ia samples (Hamuy et al.
1996b; Riess et al. 1998b) with $cz$ $\geq 2500 $ km s$^{-1}$ provides
a homogeneous training set of objects for MLCS with well understood
relative luminosities.  Here we employ a set (see Table 10) of
$B$ and $V$ light curves with $cz$ $\geq 2500 $ km s$^{-1}$ to determine
the MLCS relations.

   The significant increase in the size of the available training set
of SNe Ia since Riess, Press, \& Kirshner (1996a) supports an expansion of our
description of the MLCS relations.  Riess, Press, \& Kirshner (1996a) described SNe Ia light
curves as a {\it linear} family of the peak luminosity:
 \bq {\bf m_V}={\bf M_V}+{\bf R_V}\Delta+\mu_V \eq
 \bq {\bf m_{B-V}}={\bf M_{B-V}}+{\bf R_{B-V}}\Delta+E_{B-V} \eq
where ${\bf m_V,m_{B-V}}$ are the observed light and color curves,
$\Delta\equiv M_v-M_v(standard)$ is the
difference in maximum luminosity between the fiducial
template SN Ia and any other SN Ia, ${\bf R_V}$ and ${\bf R_{B-V}}$
 are vectors of
correlation coefficients between $\Delta$ and the light curve shape, $\mu_v$ is
the apparent distance modulus, and $E_{B-V}$ is the color excess.  All
symbols in bold denote vectors which are functions of SN Ia age, with
$t=0$ taken by convention as the epoch of $B$ maximum.  

   By adding a second-order term in the expansion, our empirical model
   becomes

\bq {\bf m_V}={\bf M_V}+{\bf R_V}\Delta+{\bf Q_V}\Delta^2+\mu_V \eq
 \bq {\bf m_{B-V}}={\bf M_{B-V}}+{\bf R_{B-V}}\Delta+{\bf Q_{B-V}}\Delta^2+E_{B-V} \eq
where ${\bf Q_V,Q_{B-V}}$ are the correlation coefficients of the
 quadratic relationship between $\Delta^2$ and the light curve shape.
The vectors of coefficients (${\bf R_V,R_{B-V},Q_V,Q_{B-V}}$) as well
 as
the fiducial templates (${\bf M_V,M_{B-V}}$) are
 determined
from the training set of SNe Ia listed in Table A.  (They can be found
 at http://oir-www.harvard.edu/cfa/oirResearch/supernova.)  The empirical
 light and color curve families are shown in Figure 13. As before,
 these MLCS relations show that the more luminous SNe Ia have broader light
 curves and are bluer until day $\sim35$, by which time all SNe Ia
 have the same color.  The primary difference from the previous MLCS
 relations is that near maximum, the color range spanned by the same
 range of SN Ia luminosities is much reduced.  Further, the quadratic
 MLCS relations reveal that SNe Ia which are brighter or dimmer (than
 the fiducial value) by equal
 amounts do not show equal changes in their colors.  Faint
 SNe Ia are far redder than the amount by which luminous SNe Ia are blue.

   Fitting of this quadratic model (equations A3-A4) to a SN Ia still
   requires the determination of 4 ``free'' parameters: $\Delta$,
   $\mu_V$, $E_{B-V}$, and $t_{max}$.  The parameters are determined by
 minimizing the expected deviations
between data and model:
\bq \chi^2={\bf r_x}C^{-1}{\bf r_x^T}, \eq
where \bq {\bf r_x}={\bf m_x}-{\bf M_x}-{\bf R_x}\Delta-{\bf
Q_x}\Delta^2-\mu_x \eq
for any band x.  Here C is the correlation matrix of the model and the
measurements.  Correlations of the data from the model were determined
from the SNe Ia of Table 10.  These correlations result from our still
imperfect (but improving) description of the light-curve shape behavior.  Future
expansion of the model will reduce these correlations further until
they become constraints on the unpredictable, turbulent behavior of
the SN Ia atmosphere.  Riess, Press, \& Kirshner (1996a) quantified the autocorrelation (diagonal
matrix elements) of the linear model.  Here we have determined, in
addition,  the covariance
(off-diagonal matrix elements) between two measurements of differing
SN Ia age, passband, or both.  These can be found at http://oir-www.harvard.edu/cfa/oir/Research/supernova.
The correlation matrix of the measurements, commonly called the
``noise,'' is, as always, provided by the conscientious observer.

The {\it a priori} values for $\Delta$ used to determine the vectors
${\bf R_V,R_{B-V},Q_V,Q_{B-V},M_V}$, and ${\bf M_{B-V}}$ are the differences between
the measured peak magnitudes and those predicted by the SN Ia host
galaxy redshift.  These values for $\Delta$ must be corrected for
the extinction, $A_V$.  Because the values of $A_V$ are not known {\it
a priori}, we use an initial guess derived from the color
excess 
measured from the uniform color range of SNe Ia after day 35 (Riess, Press, \& Kirshner
1996a; Lira 1995).  

Initial guesses for $\Delta$, $\mu_V$, $E_{B-V}$, and $t_{max}$ yield
estimates for ${\bf R_V,R_{B-V},Q_V,Q_{B-V},M_V}$, and ${\bf M_{B-V}}$ by
minimizing equation (A5) with respect to the latter.
These estimates for 
${\bf R_V,R_{B-V},Q_V,Q_{B-V},M_V}$, and ${\bf M_{B-V}}$ yield
improved estimates of $\Delta$, $\mu_V$, $E_{B-V}$, and $t_{max}$ also
determined by minimizing equation (A5) with respect
to the latter.  
This iterative determination of these vectors and parameters is
repeated until convergence is reached.  Subsequent determination of
the parameters $\Delta$, $\mu_V$, $E_{B-V}$, and $t_{max}$ for SNe
Ia not listed in Table 10 (such as those reported here) is done using the
fixed vectors derived from this training process.  

 We also employ a refined estimate of the selective absorption to
 color excess ratio, $R_V={A_V/E_{B-V}}$, which has been calculated
 explicitly as a
 function of SN Ia age from accurate spectrophotometry of SNe Ia
 (Nugent, Kim, \& Perlmutter 1998).  This work shows that although
 $R_V$ is the canonical value of  $\sim 3.1$ for SNe Ia at maximum
 light or before, over the first 10 days after maximum
 $R_V$ slowly rises to about 3.4.  For highly reddened SNe Ia, this
 change in $R_V$ over time can appreciably affect the shape of the
SN Ia light curve (Leibundgut 1989).   

   Lastly we have refined our {\it a priori} understanding of the
likelihood for SN Ia interstellar extinction from host galaxies.  The
previous incarnation of MLCS (Riess, Press, \& Kirshner 1996a) employed a ``Bayesian filter'' to combine our
measurement of extinction with our prior knowledge of its
one-directional effect.  In addition, it is less probable to observe a very large
amount of extinction due to the finite column density of a spiral disk
as well as a reduced likelihood for detection of SNe with
large extinctions.  To quantify this {\it a priori} likelihood for extinction
we have adopted the calculations of Hatano, Branch, \& Deaton (1997),
who determined the extinction distribution for SNe Ia in the bulge
and disk of late-type galaxies.  The primary difference between our
previous {\it a priori} distribution and the results of Hatano,
Branch, \& Deaton (1997)
are that non-trivial quantities of extinction are even
less probable than assumed.  In particular, Hatano, Branch, \& Deaton
(1998) show that two-thirds of SNe Ia suffer less than 0.3 to 0.5
mag of extinction, which is approximately half the amount
of extinction previously assumed.   Despite our use of an externally
derived Bayesian prior for probable SN Ia extinction, it is
important to continue testing that the {\it a posterior} extinction
distribution matches the expected one.  A statistically significant
departure could imply an important deficiency in the SN Ia luminosity, 
light-curve shape, and color relations.  Specifically, excessively
blue SNe Ia such as SN 1994D ($E_{B-V} \approx  -0.10 \pm 0.04$), 
if common, would reveal a shortcoming of these MLCS relations.  
However, using the current MLCS relations, the best estimate we can
make for such blue SNe Ia is that their extinctions are negligible.
If the {\it a priori} distributions of Hatano,
Branch, \& Deaton (1997) are not significantly in error, this practice
is statistically sensible and does not introduce a distance bias.

\vfill \eject
 
\centerline {\bf References}
\vskip 12 pt

\refitem Arnett, W.D., 1969, {\it Astrophys. Space Sci.}, 5, 280

\refitem Bahcall, J. N.,  et al. 1996, ApJ, 457,19

\refitem Bahcall, N. A., Fan, X., \& Cen, R., ApJ, 1997, 485, 53

\refitem Bolte, M. \& Hogan, C. J. 1995, Nature, 376, 399

\refitem Bouchet, P., Lequeux, J., Maurice, E., Prevot, L., \&
Prevot-Burnichon, M. L., 1985, A\&A, 149, 300

\refitem Branch, D., 1998, ARAA, in press

\refitem Branch, D., Fisher, A., \& Nugent, P., 1993, AJ, 106, 2383

\refitem Branch, D., \& van den Bergh 1993, AJ, 105, 2251

\refitem Branch, D., \& Miller, D. 1993, ApJ, 405, L5

\refitem Branch, D., \& Tammann, G.A.,1992, ARAA, 30, 359

\refitem Branch, D., et al. 1988, ApJ, 330, 117

\refitem Burstein, D., \& Heiles, C. 1982, AJ, 87, 1165

\refitem Caldwell, R. R., Dave, R. \& Steinhardt, P. J., 1998, PRL, 80, 1582

\refitem Cappellaro, E. et al. 1997, A\&A, 322, 431 

\refitem Cardelli, J.A., Clayton, G. C., \& Mathis, J. S., 1989, ApJ,
345, 245 

\refitem Carlberg, R. G., Yee, H. K. C., Ellingson, E., Abraham, R., Gravel, P., Morris, S. \& Pritchet, C. J. 1996, ApJ, 462, 32

\refitem Carroll, S. M., Press, W. H., \& Turner, E. L., 1992, ARAA, 30, 499

\refitem Chaboyer, B. 1995, ApJ, 444, L9

\refitem Chaboyer, B., Demarque, P., Kernan, P. J., \& Krauss, L. M., 1998, ApJ, 494, 96

\refitem Chaboyer, B., Demarque, P., \& Sarajedini, A. 1996, ApJ, 459,
558

\refitem Clocchiatti, A. et al. 1998, in preparation

\refitem Colgate, S. 1979, ApJ, 232, 404

\refitem Colgate, S., \& McKee, C. 1969, ApJ, 157, 623

\refitem Cowan, J. J., McWilliam, A., Sneden, C., \& Burris,
D. L. 1997, ApJ, 480, 246

\refitem Cristiani, S., et al. 1992, A\&A, 259, 63

\refitem Della Valle, M., \& Panagia, N. 1992, AJ, 104,696

\refitem Doggett, J. B., \& Branch, D. 1985, AJ, 90, 2303

\refitem Feast, M. W., \& Catchpole, R. M. 1997,  \mnras, 286, 1p 

\refitem Feast, M. W., \& Walker, A. R. 1987, ARAA, 25, 345

\refitem Filippenko, A.V. 1997, ARA\&A, 35, 309

\refitem Filippenko, A.V. et al. 1992a, AJ, 384, 15

\refitem Filippenko, A.V. et al. 1992b, AJ, 104, 1543

\refitem Filippenko, A. V., et al. 1998, in preparation

\refitem Ford, C. et al. 1993, AJ, 106, 1101

\refitem Fukugita, M., Futamase, T., \& Kasai, M., 1990, MNRAS, 246, 24

\refitem Garnavich, P., et al. 1996a, IAUC 6332

\refitem Garnavich, P., et al. 1996b, IAUC 6358

\refitem Garnavich, P., et al. 1996c, IAUC 6861

\refitem Garnavich, P., et al. 1998, ApJ, 493, 53

\refitem Garnavich, P., et al. 1998, in preparation

\refitem Goobar, A. \& Perlmutter, S. 1995, ApJ, 450, 14

\refitem Gratton, R. G., Fusi Pecci, F., Carretta, E.,
Clementini, G., Corsi, C. E., \& Lattanzi, M. 1997, ApJ,
491, 749

\refitem Hamuy, M., \& Pinto, P. 1998, in preparation

\refitem Hamuy, M., Phillips, M. M., Suntzeff, N. B., Schommer, R. A., 
Maza, J., \& Avil\'es, R. 1996a,
AJ, 112, 2398 

\refitem Hamuy, M., et al. 1996b,
AJ, 112, 2408

\refitem Hamuy, M., Phillips, M. M., Maza, J., Suntzeff, N. B., Schommer, 
R. A., \& Avil\'es, R. 1995, AJ, 109, 1

\refitem Hamuy, M., Phillips, M. M., Schommer, R. A., Suntzeff, N. B.,  
Maza, J., \& Avil\'es, R. 1996c,
AJ, 112, 2391

 \refitem Hamuy, M., Phillips, M. M., Suntzeff, N. B., Schommer, R. A., 
Maza, J., Smith, R. C., Lira, P., \& Avil\'es, R. 1996d,
AJ, 112, 2438 
 
\refitem Hamuy, M., et al. 1994, AJ, 108, 2226

\refitem Hamuy, M., et al. 1993a, AJ, 106, 2392

\refitem Hamuy, M., Phillips, M. M., Wells, L. A., \& Maza, J. 1993b,
PASP, 105, 787

\refitem Hansen, L., Jorgensen, H. E., \& N\o rgaard-Nielsen, H. U., 1987, ESO Msngr, 47, 46

\refitem Hatano, K., Branch, D., \& Deaton, J., 1997, astro-ph/9711311

\refitem Hodge, P. W., \& Kennicutt, R. C. 1982, 87, 264

\refitem H\"{o}flich, P., Thielemann, F. K., \& Wheeler, J. C. 1998, ApJ, 495, 617

\refitem Holz, D. E. 1998, private communication

\refitem Holz, D. E., \& Wald, R. M. 1998, submitted (astro-ph/9708036)

\refitem Hoyle, F., \& Fowler, W.A., 1960, ApJ, 132, 565

\refitem Huang, J. S., Cowie, L. L., Gardner, J. P., Hu, E. M., Songaila, A., \& Wainscoat, R. J., 1997, ApJ, 476, 12

\refitem Kantowski, R., Vaughan, T., \& Branch, D. 1995, ApJ, 447, 35

\refitem Kim, A., Goobar, A., \& Perlmutter, S. 1996, PASP, 108, 190

\refitem Kim, A., et al. 1997, ApJ, 476, 63

\refitem Kirshner, R. P., et al. 1995, IAUC No. 6267

\refitem Kochanek, C. S. 1996, ApJ, 466, 638

\refitem Kochanek, C. S. 1997, ApJ, 491, 13

\refitem Kowal, C. T. 1968, AJ, 73, 1021

\refitem Landolt, A. U. 1992, AJ, 104, 340

\refitem Leibundgut, B. et al. 1993, AJ, 105, 301

\refitem Lin, H., et al. 1996, ApJ, 471, 617

\refitem Lira, P., 1995, Masters thesis, Univ. of Chile

\refitem Lupton, R., 1993, ``Statistics in Theory and Practice,''
Princeton University Press, Princeton, New Jersey

\refitem Mateo, M., \& Schechter, P. L. 1989,, in Proc. 1st ESO/ST-ECF
Workshop on Data Analysis, ed. P. J. Grosbol, F. Murtaugh, \&
R. H. Warmels (Garching: ESO)

\refitem Madore, B. F. \& Freedman, W. L., 1998, ApJ, 492, 110

\refitem Madore, B. F. \& Freedman, W. L., 1991, PASP, 103, 933

\refitem  Marzke, R.O., Geller, M.J., daCosta, L.N., \& Huchra,
J.P. 1995, AJ, 110, 477

\refitem Menzies, J. W. 1989, MNRAS, 237, 21

\refitem Miller, D. \& Branch, D. 1990, AJ, 100, 530

\refitem N\o rgaard-Nielsen, H., et al. 1989, Nature, 339, 523

\refitem Nugent, P., Phillips, M., Baron, E., Branch, D., \&
Hauschildt, P. 1995, ApJ, 455, L147

\refitem Nugent, P., Kim, A., \& Perlmutter, S. 1998, in preparation

\refitem Nugent, P., et al. 1998b, in preparation

\refitem Nugent, P., et al 1998a, IAUC 6804

\refitem Oke, J. B., \& Sandage, A. 1968, ApJ, 154, 21

\refitem Oke, J. B., et al. 1995, PASP, 107, 375

\refitem Oswalt, T. D., Smith, J. A., Wood, M. A., \& Hintzen, P. 1996, Nature,
382, 692

\refitem Perlmutter, S., et al., 1995, ApJ, 440, 41

\refitem Perlmutter, S., et al., 1997, ApJ, 483, 565

\refitem Perlmutter, S., et al., 1998, Nature, 391, 51

\refitem Phillips, M. M. 1993, ApJ, L105, 413

\refitem Phillips, M. M., Wells, L., Suntzeff, N., Hamuy, M., Leibundgut, B., Kirshner, R.P., \& Foltz, C. 1992, AJ, 103, 1632

\refitem Phillips, M. M., et al. 1987, PASP, 99, 592

\refitem Phillips, M. M. et al 1998, in preparation

\refitem Pierce, M. J., 1994, ApJ, 430, 53

\refitem Porter, A. C., \& Filippenko, A. V. 1987, AJ, 93, 1372

\refitem Pskovskii, Y. 1984, {\it Sov. Astron.}, 28, 658

\refitem Reid, I. N. 1997, AJ, 114, 161

\refitem Reiss, D. J., Germany, L. M., Schmidt, B. P., \& Stubbs, C. W. 1998, AJ, 115, 26

\refitem Riess, A. G. 1996, PhD thesis, Harvard University

\refitem Riess, A. G., Press W.H., \& Kirshner, R.P., 1995, ApJ, 438 L17 

\refitem Riess, A. G., Press, W.H., \& Kirshner,  R.P. 1996a, ApJ, 473,
88 

\refitem Riess, A. G., Press, W.H., \& Kirshner,  R.P. 1996b, ApJ, 473,
588 
 
\refitem Riess, A. G., et al. 1997, AJ, 114, 722

\refitem Riess, A. G., Nugent, P. E., Filippenko, A. V., Kirshner, R. P., \& Perlmutter, S., 1998a, ApJ, in press

\refitem Riess, A. G., et al. 1998b, in preparation 

\refitem Renzini, A., Bragaglia, A., Ferraro, F. R., Gilmozzi, R., Ortolani, S., Holberg, J. B., Liebert, J., Wesemael,
F., \& Bohlin, R. C. 1996, ApJ, 465, L23

\refitem Rood, H.J., 1994, PASP, 106, 170

\refitem Ruiz-Lapuente, P., Cappellaro, E.,
 Turatto, M., Gouiffes, C.,
 Danziger, I. J.,
 Della Valle, M., \& Lucy, L. B. 1992, ApJ, 387, 33

\refitem Rybicki, G. B., \& Press, W. H. 1992, ApJ, 398, 169

\refitem Sandage, A., \& Tammann, G. A. 1993, ApJ, 415, 1    

\refitem Sandage, A., et al. 1994, ApJ, 423, L13

\refitem Saha, A., et al. 1994, ApJ, 425, 14

\refitem Saha, A., et al. 1997, ApJ, 486, 1

\refitem Savage, B. D., \& Mathis, J. S., 1979, ARAA, 17, 73

\refitem Schechter, P. L., Mateo, M., \& Saha, A. 1993, PASP, 105, 1342

\refitem Schmidt, B. P, et al. 1998, ApJ, in press

\refitem Strauss, M.A., \& Willick, J.A., 1995, PhR, 261, 271
  
\refitem Tammann, G.A., \& Leibundgut, B. 1990, A\&A, 236, 9

\refitem Tonry, J.L., 1991, ApJ, 373, L1

\refitem Turatto, M., Benetti, S., Cappellaro, E., Danziger, I. J., Della Valle, M., 
     Gouiffes, C., Mazzali, P. A., \& Patat, F. 1996, MNRAS, 283, 1

\refitem Turner, E. L., 1990, ApJ, 365, 43

\refitem Uomoto, A., \& Kirshner, R.P., 1985, A\&A, 149, L7

\refitem Vandenberg, D. A., Stetson, P. B., \& Bolte, M., 1996, ARAA 34, 461

\refitem von Hippel, T., Bothun, G. D., \& Schommer, R. A. 1997, AJ,
114, 1154

\refitem Walterbos, R. 1986, PhD Thesis, Leiden University

\refitem Wambsganss, J., Cen, R., Guohong, X., \& Ostriker, J. 1997, ApJ, 475, L81

\refitem Wheeler, J. C., \& Levreault, R. 1985, ApJ, 294, L17

\refitem Wheeler, J. C., \& Harkness, R.P., 1990, Rep. Prog. Phys., 53, 1467
  
\refitem Wheeler, J. C., Harkness, R.P., Barkat, Z., \& Swartz, D. 1986, PASP, 98, 1018

\refitem White, M., \& Scott, D. 1996, ApJ, 459, 415.

\refitem Zehavi, I., Riess, A. G., Kirshner, R. P., \& Dekel, A. 1998, ApJ, in press

\vfill \eject
 
{\bf Figure 1:}  Identification spectra (in f$_{\lambda}$) of high-redshift SNe Ia.  The spectra
obtained for the 10 new SNe of the high-redshift sample are shown in
the restframe.  The data are
compared to nearby SN Ia spectra of the same age as determined by the
light curves (see Table 1).  The spectra the three objects from
Garnavich et al. (1998) are also displayed.

{\bf Figure 2:}  Local standard stars in the fields of SNe Ia.
The stars are listed
in Table 2 and the locations of the stars and SNe are indicated in the figure.  The orientation of each field
is East to the right and North at the top.  The width and length of
each field is: 96E=4.9$^\prime$, 96H=4.9$^\prime$, 96I=4.9$^\prime$, 96J=4.9$^\prime$, 96K=4.9$^\prime$,
96R=5.0$^\prime$, 96T=4.9$^\prime$, 96U=4.9$^\prime$, 95ao=4.8$^\prime$, 95ap=4.8$^\prime$.

{\bf Figure 3:} Light curves of high-redshift SNe Ia.  $B$ (filled
symbols) and $V$ (open symbols) photometry in the rest-frame of 10
well-observed SNe Ia is shown with $B$ increased by 1 mag for ease
of view.  The lines are the empirical MLCS model fits to the
data.  Supernova age is shown relative to $B$ maximum.

{\bf Figure 4:} MLCS SNe Ia Hubble diagram.  The upper panel shows the
Hubble diagram for the low-redshift and high-redshift SNe Ia samples
with distances measured from the MLCS method (Riess, Press, \& Kirshner 1995, 1996a; Appendix
of this paper).  Overplotted are three cosmologies: ``low'' and
``high'' $\Omega_M$ with $\Omega_\Lambda=0$ and the best fit
for a flat cosmology, $\Omega_M=0.24$, $\Omega_\Lambda=0.76$.  The bottom panel
shows the difference between data and models with $\Omega_M=0.20$,
$\Omega_\Lambda=0$.  The open symbol is SN 1997ck
($z=0.97$) which
lacks spectroscopic classification and a color measurement.  The
average difference between the data and the $\Omega_M=0.20$,
$\Omega_\Lambda=0$ prediction is 0.25 mag.

{\bf Figure 5:} $\Delta m_{15}(B)$ SN Ia Hubble diagram.  The upper panel shows the
Hubble diagram for the low-redshift and high-redshift SNe Ia samples
with distances measured from the template fitting method parameterized
by $\Delta m_{15}(B)$ (Hamuy et al. 1995, 1996d).  Overplotted are three cosmologies: ``low'' and
``high'' $\Omega_M$ with $\Omega_\Lambda=0$ and the best fit for a flat
cosmology, $\Omega_M=0.20$, $\Omega_\Lambda=0.80$.  The bottom panel
shows the difference between data and models from the $\Omega_M=0.20$,
$\Omega_\Lambda=0$ prediction. The open symbol is SN 1997ck
($z=0.97$) which
lacks spectroscopic classification and a color measurement.  The
average difference between the data and the $\Omega_M=0.20$,
$\Omega_\Lambda=0$ prediction is 0.28 mag.

{\bf Figure 6:} Joint confidence intervals for ($\Omega_M$,$\Omega_\Lambda$)
from SNe Ia.  The solid contours are results from the MLCS method
applied to well-observed SNe Ia light curves together with the
snapshot method (Riess et al. 1998a) applied to incomplete SNe Ia light
curves.  The dotted contours are for the same objects excluding the
unclassified SN 1997ck ($z=0.97$).   Regions
representing specific cosmological scenarios are illustrated.
Contours are closed by their intersection with the line $\Omega_M=0$.

{\bf Figure 7:} Joint confidence intervals for ($\Omega_M$,$\Omega_\Lambda$)
from SNe Ia.  The solid contours are results from the template fitting method
applied to well-observed SNe Ia light curves together with the
snapshot method (Riess et al. 1998a) applied to incomplete SNe Ia light
curves. The dotted contours are for the same objects excluding the
unclassified SN 1997ck ($z=0.97$).   Regions
representing specific cosmological scenarios are illustrated.
Contours are closed
 by their intersection with the line $\Omega_M=0$.

{\bf Figure 8:} PDF for the dynamical age of the Universe from SNe
Ia (equation 19). The PDF for the dynamical age derived
from the PDFs for $H_0$, $\Omega_M$,$\Omega_\Lambda$
is shown for the two different distance methods without the unclassified
SN 1997ck.  A naive average (see \S 4.2) yields an
estimate of 14.2$^{+1.0}_{-0.8}$ Gyr, not including the systematic
uncertainties in the Cepheid distance scale.

{\bf Figure 9:} Lines of constant dynamical age in Gyr in the
($\Omega_M$,$\Omega_\Lambda$) plane.  Comparing these lines with the
error ellipses in Figures 5 and 7 reveals the leverage this experiment
has on measuring the dynamical age. This plot assumes $H_0=65$ km
s$^{-1}$ Mpc$^{-1}$ as determined from nearby SNe Ia and is subject to
the zeropoint of the Cepheid distance scale.

{\bf Figure 10:} Distributions of MLCS light curve shape parameters,
$\Delta$, and template fitting parameters, $\Delta m_{15}(B)$, for the high and low-redshift samples of SNe Ia.  Positive
values for $\Delta$ and $\Delta m_{15}(B) > 1.1$ correspond to intrinsically dim SNe Ia, negative
values for $\Delta$ and $\Delta m_{15}(B) < 1.1 $ correspond to luminous SNe Ia.  Histograms
of the low-redshift (solid line) and high-redshift (dotted line) light
curve shape parameters are mutually consistent with no indication that
these samples
are drawn from different populations of SNe Ia.  Filled and open circles
show the distribution of log($cz$) for the low and high-redshift
samples, respectively.

{\bf Figure 11:} Spectral comparison (in f$_{\lambda}$) of SN 1998ai ($z=0.49$) with
low-redshift ($z < 0.1$) SNe Ia at a similar age.  Within the narrow range
of SN Ia spectral features, SN 1998ai is indistinguishable
from the low-redshift SNe Ia.  The spectra from top to bottom
are SN 1992A, SN 1994B, SN 1995E, SN 1998ai, and SN 1989B  
$\sim$ 5 days before maximum light.  The spectra of the low-redshift
SNe Ia were resampled and convolved with Gaussian noise to match
the quality of the spectrum of SN 1998ai.  

{\bf Figure 12:} Comparison of the spectral and photometric
observations of SN 1996E to those of Type Ia and Type Ic supernovae.  The low
signal-to-noise ratio of the spectrum of SN 1996E and the absence of
data blueward of 4500 \AA\ makes it difficult to distinguish
 between a Type Ia and Ic classification.  The light and color curves of SN 1996E are also
consistent with either supernova type.  The spectrum was taken six
days (rest-frame) after the first photometric observation.

{\bf Figure 13:} MLCS empirical SN Ia light curve families in $M_B$, $M_V$,
and $(B-V)_0$.  The derived light curves are given as a function of
the luminosity difference, $\Delta$, between the peak visual
luminosity of a SN Ia and a fiducial ($\Delta=0$) SN Ia.  Properties
of the SN Ia families
 are indicated in the figure and the Appendix.  The light and color curves of SN 1995ac (open symbols) and SN 1996X (filled symbols) are overplotted as examples of luminous and dim SN Ia, respectively.

\vfill \eject

\end{document}